\documentclass[useAMS]{mn2e}
\usepackage{epsfig,graphics}
\usepackage{amsmath}
\usepackage{times}
\usepackage{subfig}
\usepackage{amsmath, amssymb, graphicx, multirow, rotating, lscape}
\usepackage{float}
\DeclareCaptionFormat{cont}{#1 (cont.)#2#3\par}
\title[Spectral evolution of 3C 273]
{Origin of X-rays in the low state of the FSRQ 3C 273: Evidence of inverse Compton emission}

\author[Kalita et al.]{Nibedita\ Kalita$^{1,2}$\thanks{Email: nibeditaklt1@gmail.com}, 
Alok C.\ Gupta$^{3,1}$\thanks{Email: acgupta30@gmail.com}\thanks{PIFI Visiting Scientist}, 
Paul J.\ Wiita$^{4}$\thanks{Email: wiitap@tcnj.edu}
Gulab C.\ Dewangan$^{5}$, 
\newauthor Kalpana\ Duorah$^{2}$ \\
\\      
\\
$^1$Aryabhatta Research Institute of Observational Sciences (ARIES), Manora Peak, Nainital, 263 002, India \\
$^2$Department of Physics, Gauhati University, Guwahati, 781 014, India  \\   
$^3$Key Laboratory for Research in Galaxies and Cosmology, Shanghai Astronomical Observatory, Chinese Academy 
of Sciences, 80 Nandan Road, \\
~~Shanghai 200030, China \\
$^4$Department of Physics, The College of New Jersey, P.O.\ Box 7718, Ewing, NJ 08628-0718, USA  \\     
$^5$Inter University Centre for Astronomy and Astrophysics (IUCAA), Post Bag 4, Pune 411 007, India \\              
}

\begin{document}
\date{Accepted 2017 May 5. Received 2017 May 4; in original form 2016 August 28}

\maketitle
\label{firstpage}

\pagerange{\pageref{firstpage}--\pageref{lastpage}} \pubyear{2016}

\begin{abstract}

We analyze the 2.5--10 keV X-ray spectra of the luminous quasar 3C 273 and simultaneous observations in UV wavelengths from {\it XMM--Newton} between 2000 and 2015.   The lowest  flux level ever was  observed in 2015.
The continuum emission from 3C 273 is generally best described by an absorbed power-law but during extremely low states the addition of fluorescence from the K-shell iron line improves the fit. We study the spectral evolution of the source during its extended quiescent state  and also examine connections between the X-ray and ultraviolet emissions, which have been seen in some, but not all, previous work.
We detect a possible anti-correlation between these two bands during the low state that characterized 3C 273 for most of this period; however, this was not present during a flaring state. 
 A harder-when-brighter trend for the X-ray spectrum was observed in these long-term observations of 3C 273 for the first time.  We suggest that the X-ray emission in 3C 273 is the result of inverse Compton scattering of soft UV seed photons (emitted from the local environment of the AGN), most likely in a thermal corona. We can explain the significant temporal variation of the spectral continuum  as an outcome of changing optical depth of the comptonizing medium, along the lines of the wind-shock model proposed by Courvoisier and Camenzind (1989). 
\end{abstract}


\section{Introduction}

The first discovered quasar, 3C 273 (Schmidt 1963), is classified as a Flat Spectrum Radio Quasar (FSRQ)  and thus belongs to the violently variable AGN class of blazars. This quasar is  nearby ($z = 0.158339$) and luminous in all frequencies. It shows most of the common properties of blazars, including high polarization and large flux variability at optical wavebands, strong radio emission, and a relativistically outflowing nonthermal radio jet oriented close to our line of sight. Radio through infrared/optical emissions are dominated by synchrotron emission from the jet (Robson et al.\ 1993; T{\"u}rler et al.\ 2000). Moreover, the object contains a radio core which has been found to show superluminal motions (Jorstad et al.\ 2001, 2005).

X-ray emission of 3C 273 possesses some peculiar features, which are generally not common in blazars. A variable soft component below 2 keV is found in the source in almost all satellite observations (e.g. Turner et al.\ 1985; Courvoisier et al.\ 1987; Leach et al. 1995; Page et al.\ 2004; T{\"u}rler et al.\ 2006; Chernyakova et al.\ 2007; Pietrini \& Torricelli-Ciamponi\ 2008;  and references therein) which is generally fitted with multiple black body components or a Comptonization model (Page et al.\ 2004) or even with a steep power-law. The presence of a Fe K$\alpha$ emission line near 6.4 keV has been detected from time to time by several authors (Turner et al.\ 1990; Yaqoob \& Serlemitsos 2000; Page et al.\ 2004). A large optical/ultraviolet bump (Big Blue Bump -- BBB), which can be explained as result of accretion disk emission (Paltani et al.\ 1998), is another interesting feature generally found in this source. Surprisingly,  all these features are found in Seyfert galaxies, which are closer and weaker AGN as compared to blazars; moreover, the angle  between the jet direction (when present) and our line of sight is much larger for Seyferts than for blazars (Urry \& Padovani 1995). The presence of both a BBB and soft excess in the source indicates a possible link between these two emission phenomena. Regarding all these characteristics it can be said that X-ray emission from 3C 273 is a combination of blazar-like and Seyfert-like components. To explore these questions and obtain a better understanding of the physics, we need to study the X-ray spectra of 3C 273 in different states. 

3C 273 was first detected in the 1--10 keV X-ray band by Bowyer et al.\ (1970) and since then it has been observed and studied extensively with all the major X-ray satellites. During bright states, the 2--10 keV spectra were fit by a power-law with spectral index of around 1.5, while the 10--25 keV band could be fit with a power-law of spectral index 0.91 using {\it EXOSAT} observations (Turner et al.\ 1985). This confirms that the high energy X-ray spectrum of 3C 273 has a hard tail when the source is bright (Turner et al.\ 1985). With 5 years of {\it EXOSAT} and  {\it Ginga} observations, Turner et al.\ (1990) reported that the X-ray spectrum is well described by a power-law with Galactic absorption and that the power-law spectral index is not correlated to the 2--10 keV flux. The absence of correlation between photon index and flux was also reported by Page et al.\ (2004). It was also discovered that this power-law spectrum  may extend up to 200 keV (0.1--200 keV) using {\it BeppoSAX} data (Haardt et al.\ 1998), with a spectral index of $\Gamma = 1.53 - 1.6$.  The occasional appearance of a weak iron line, either narrow or broad, has been found at an energy of nearly 6.4 keV (Haardt et al.\ 1998; Yaqoob \& Serlemitsos 2000; Page et al.\ 2004). In a few cases, a more complicated model, involving a cold reflector in addition to power-law and black body components, such as  {\it PEXRAV}, seemed to be required, as  when a deviation from power-law spectra below 1 keV and above 10 keV was reported with {\it BeppoSAX} observations during 1996--2001 (Grandi $\&$ Palumbo 2004). Kataoka et al.\ (2002) found that the source flux varies by a factor of 4 over a period of four years and the hard power-law spectrum is not due to the contribution of Seyfert-like beamed emission but likely arose from inverse Compton scattering emitted by the jet. So, generally it can be said that the harder continuum in the energy range 2--10 keV of 3C 273 is well described by an absorbed power-law with a differential spectral index.

Specific features in the X-ray spectra of 3C 273, i.e., the fluorescent $K{\alpha}$ transition line, the absorption edge and the soft excess, are found to be variable in different observations taken at different times (Haardt et al.\ 1998), and so these variations were interpreted as being related to the state of the source. During the source low state the contributions from thermal and non-thermal emission become comparable, hence the X-ray features become prominent.  At higher luminosity states, emission from the relativistic jet dominates any thermal and/or reprocessed emission related to the disc, and those features are diluted. Clearly the X-ray spectra of 3C 273 are not as simple as those of other blazars. Several emission components contribute to the X-ray spectra, resulting in appearance of distinct features which  depend on the different states of the source.

Unlike lower frequency emissions, the X-rays  can not be fully explained by the synchrotron emission mechanism, but there is debate as to whether the excess predominantly results from inverse Compton scattering of low energy seed photons by the relativistic electrons in the jet or by the hot corona above the accretion disc (Haardt $\&$ Maraschi 1991).  There are no strong results that   fully support either explanation.  A common origin of the UV and X-ray emissions was argued for by Walter $\&$ Courvoisier (1992). However, later observations were unable to detect a connection between these two bands (Chernyakova et al.\ 2007; Soldi et al.\ 2008), which makes the contributions of different X-ray  emission mechanisms to the flux quite uncertain. 

Observations made by {\it XMM-Newton} have been used to study the spectral behaviour of 3C 273 in different epochs. The observations made during 2000--2003 were studied by Page et al.\ (2004) who found that the spectra below 2 keV was dominated by a strong soft excess which can be adequately fitted by either multiple black bodies or a Comptonization model and the 3--10 keV spectra were well fitted by a  power-law model with an occasional weak broad Fe emission line (EW $\sim$ 56 eV).  Spectral study of the source during 2000--2004 with {\it XMM--Newton} revealed an increase of the thermal component as compared to the synchrotron component (producing a softening of the hard spectral index) where the 0.4--10 keV spectra were described by broken power-law model (Foschini et al.\ 2006). Courvoisier et al.\ (2003) used only the spectra above 3 keV, avoiding the soft excess, to study the cross calibration between different satellite's instruments, where all the spectra were fitted with a simple power-law model.  The presence of a quasi-periodicity (QPO) of period 3.3 ks in 3C 273 (during Obs ID 1267003013 in 2000) was claimed (Espaillat et al.\ 2008) but this was later deemed to be not significant by Gonz{\'a}lez-Mart{\'{\i}}n \& Vaughan (2012).  

The broadband spectrum (0.2--100 keV) of the object during 2003 to 2005 made with quasi-simultaneous {\it INTEGRAL} and {\it XMM--Newton} observations was well described by a soft cut-off power-law plus a hard power-law (Chernyakova et al.\ 2007). They found that replacement of the soft cut-off power-law with either a black body model or by a disk reflection model did not improve the fit. They also reported the source was evolving toward softer X-ray emission over the last 30 years. Energy dependent variability of 3C 273 was reported by Soldi at al.\ (2008) in their multi-wavelength study of the source. They proposed the presence of two X-ray emitting components, where they argued that the hard X-ray emission is not related to the acceleration of electrons by the shocks in the jet, but rather is associated with the long term optical variation and emitted through inverse Compton emission (synchrotron self Compton -- SSC and/or external Compton -- EC). 
 A multi-wavelength variability study of the source using 12 years of {\it XMM-Newton} observations (2000--2012), found that the source was highly variable in all bands from optical to X-rays on long time scales but that there was no evidence for  intraday variability in the individual light curves in X-rays (Kalita et al.\ 2015). We also noticed that the amount of X-ray variability disagreed with the frequency dependent variability that is generally found in blazars. The presence of both long term variability and the lack of frequency dependent variability in the X-ray bands of 3C 273 leads us to expect interesting spectral evolution of the source during the observational period. 

The quasi-continuous observations during 2005--2012 made 
by {\it RXTE-PCA} provide us with the necessary information about the emission state of the source in X-ray band during that period (Esposito et al.\ 2015). Gathering all this information we can say that all the {\it XMM-Newton} pointings included in this study were taken during low states of 3C 273 except for the observation taken in December 2007, when an X-ray flare was detected  (Esposito et al.\ 2015). 

Only the observations taken during 2000 to 2005 by {\it XMM-Newton} (discussed above) have been extensively used in previous spectral studies of 3C 273. The observations made after 2005 have not been used to provide a more complete sample for  this purpose. Moreover, most of the previous X-ray spectral studies of 3C 273 were carried out for relatively short periods (a few years). In this work we have undertaken analyses of 15 years of observations, which makes this the longest data set available from {\it XMM-Newton} employed to study the spectral evolution of the source. We have analyzed a total 20 X-ray epochs from 2000 to 2015, with 14 of those accompanied by truly simultaneous UV/optical observations. The observations made during 2013 to 2015 are presented for the first time in this work and all of the data has benefited from being analyzed in a uniform fashion with updated pipelines. 

A broad-band X-ray study of 3C 273 was made by Madsen et al.\ (2015) with observations taken by {\it Chandra, INTEGRAL, Suzaku, Swift, XMM--Newton and NuSTAR};   the overall spectrum obtained by combining all the data except from {\it INTEGRAL} does not follow a single power-law, showing significant deviation in the energy range 30--78 keV. They found that the 3--78 keV spectrum obtained from {\it NuSTAR} is an exponentially cutoff power-law showing a weak reflection component from cold, dense material (using the pexrav model). Jet emission is apparently also observed over 30--40 keV while adding both {\it NuSTAR and INTEGRAL} data, as it requires an additional power-law for the jet component with photon index of 1.05$\pm$0.4. Another contemporaneous broadband study with X-ray and gamma-ray measurements indicated a two-component model, where the X-ray emission likely originates from the Seyfert component while the $\gamma$-ray flux is dominated by jet emission (Esposito et al.\ 2015). The correlated emission they found between radio and $\gamma$-ray bands indicates a common origin through synchrotron mechanism in the frame of a shock-in-jet model, while there was no relation between radio and X-ray bands. To fit the 1 keV to 10 GeV SED they used a cutoff power-law for the lower energy band and a log-parabola for higher energies to model the jet component. Spectral variability between different flaring states are most likely governed by change in electron distribution (Esposito et al. 2015).

An important finding of this work, that the origins of X-ray and UV/optical emissions from 3C 273 are closely related, which we interpret through inverse Compton process (see Section 4.2 and 5 for details), follows previous research by Walter $\&$ Courvoisier (1992) and Chernyakova et al.\ (2007). In the former paper, the authors found a correlation between X-ray and UV bands, using 5 years of quasi-simultaneous UV and X-ray observations (with {\it EXOSAT} and {\it Ginga}) and this supported the Comptonization model. However, the latter paper could not find the presence of any relation between these bands with simultaneous {\it XMM-Newton} observations in the period 2003--2005. This produced a puzzle for our understanding about the physical mechanisms responsible for X-ray emission in the FSRQ and led us to investigate how the UV-X-ray connection varies in the source on different timescales. We obtain an excellent correlation between these bands, which is presented in Section 4 and discussed in the following section.  Crucially, this correlation seems to be present only when 3C 273 was in low state. During  a flaring state the X-ray and UV/optical emissions behave differently and do not follow the relation found in low states. How these bands are connected in high states is still an open question, and will require further observations to answer.

In this paper we present all the simultaneous observations of 3C 273 with the EPIC-pn and Optical Monitor (OM) instruments on board the {\it XMM--Newton} satellite. Considering the large number and high quality EPIC-pn spectra over a very long time span (mid-2000 through mid-2015)  we can productively study the spectral evolution over these years.  Flux variations of the hard component of 3C 273 (here ``hard" is defined as 2.5--10 keV, following Pietrini \& Torricelli-Ciamponi 2008)  allow us to investigate correlations between different spectral parameters. We have also studied the relation of this hard continuum with UV emission using simultaneous UV and X-ray observations for a better understanding of the long term behaviour of the object. The results of soft X-ray spectra (0.6-2.5 keV), mainly focused on the soft excess measured with RGS and EPIC-pn exposures, will be presented in a future paper (Kalita et al., in preparation). {\it XMM-Newton} observations and our spectral analysis procedure are described in section 2.  Section 3 discusses the software and method used for spectral fitting. In section 4 we give our results, a discussion follows in section 5, and our conclusions are given in section 6.   

\begin{table*}

{\bf Table 1. Log of {\it XMM-Newton} on-axis observations of 3C 273 having data length $\geq$ 8 ks}\\
\small

\begin{tabular}{lccccccccc} \hline \hline
 Date of Obs.  & Obs.ID     &Revolution &Window     & GTI    &Pile up &Filter     & $\mu$            &OM filter$^1$&Exposures$^2$    \\
 dd.mm.yyyy    &            &           &Mode       &(ks)    &        &           &(counts $s^{-1}$) &             &    \\\hline
 
 13.06.2000    & 0126700301 & ~~94        &Small      & 64.9   & No       & MEDIUM    & 47.04$\pm$0.81   & 1,2,3,4,5,6  & 29    \\
 15.06.2000    & 0126700601 & ~~95        &Small      & 29.6   & No       & MEDIUM    & 45.41$\pm$0.80   & 4,5,6        & 15    \\
 15.06.2000    & 0126700701 & ~~95        &Small      & 29.9   & No       & MEDIUM    & 44.16$\pm$0.80   & 1,2,3        & 15   \\
 17.06.2000    & 0126700801 & ~~96        &Small      & 60.1   & No       & MEDIUM    & 44.26$\pm$0.79   & 1,2,3,4,5,6  & 30    \\
 13.06.2001    & 0136550101 & ~277       &Small      & 88.5   & Moderate & MEDIUM    & 21.58$\pm$0.58   & 1,2,3,4,5,6  & 12   \\
 05.01.2003    & 0136550501 & ~563       &Small      &  ~8.4   & No       & MEDIUM    & 63.22$\pm$0.94   & 2,3          & ~2    \\
 07.07.2003    & 0159960101 & ~655       &Small      & 58.0   & Weak     & THIN1     & 26.05$\pm$0.57   & 1,3,4,5,6    & ~5 \\
 08.07.2003    & 0112770501 & ~655       &Small      &  ~8.0   & No       & THIN1     & 70.66$\pm$1.04   & Nil          & Nil  \\
 14.12.2003    & 0112771101 & ~735       &Small      &  ~8.3   & No       & THIN1     & 53.34$\pm$0.87   & Nil          & Nil  \\
 30.06.2004    & 0136550801 & ~835       &Small      & 19.7   & No       & MEDIUM    & 45.27$\pm$0.81   & 1,2,3,4,5,6  & ~6  \\
 10.07.2005    & 0136551001 & 1023      &Small      & 27.5   & No       & MEDIUM    & 49.68$\pm$0.84   & 1,2,3,4,5,6  & ~6    \\
 12.01.2007    & 0414190101 & 1299      &Small      & 76.5   & Moderate & MEDIUM    & 21.42$\pm$0.52   & 1,2,3,4,5,6  & ~9    \\
 25.06.2007    & 0414190301 & 1381      &Small      & 31.9   & No       & MEDIUM    & 45.90$\pm$0.80   & 1,2,3,4,5,6  & ~8    \\
 08.12.2007    & 0414190401 & 1465      &Small      & 35.3   & Weak     & MEDIUM    & 38.16$\pm$0.69   & 1,2,3,4,5,6  & ~6    \\
 09.12.2008    & 0414190501 & 1649      &Small      & 40.4   & Weak     & THIN1     & 21.57$\pm$0.52   & 1,2,3,4,5,6  & ~6   \\
 20.12.2009    & 0414190601 & 1837      &Small      & 31.3   & Weak     & THIN1     & 24.36$\pm$0.57   & Nil          & Nil \\
 10.12.2010    & 0414190701 & 2015      &Small      & 35.8   & Weak     & THIN1     & 18.29$\pm$0.48   & 1,2,3,4,5,6  & ~6  \\
 12.12.2011    & 0414190801 & 2199      &Small      & 42.8   & No       & THIN1     & 47.48$\pm$0.82   & 1,2,3,4,5,6  & ~6   \\
 16.07.2012    & 0414191001 & 2308      &Small      & 25.5   & No       & THIN1     & 41.22$\pm$0.93   & 4,5,6        & ~3  \\
 13.07.2015    & 0414191101 & 2856      &Small      & 70.8   & Moderate & THIN1     & 21.92$\pm$0.56   & 1,2,3,4,5,6  & ~6  \\\hline
\end{tabular}     \\
$\mu$ = mean count rate in the energy range 0.3--10 keV; GTI = Good Time Interval, $^1$ 1 = {\it UVW2}, 2 = {\it UVM2}, 3 = {\it UVW1},\\ 4 = {\it U}, 5 ={\it  B}, 6 = {\it V}, $^2$ Total number of exposures combining all six filters \\
\end{table*}

\begin{figure*}
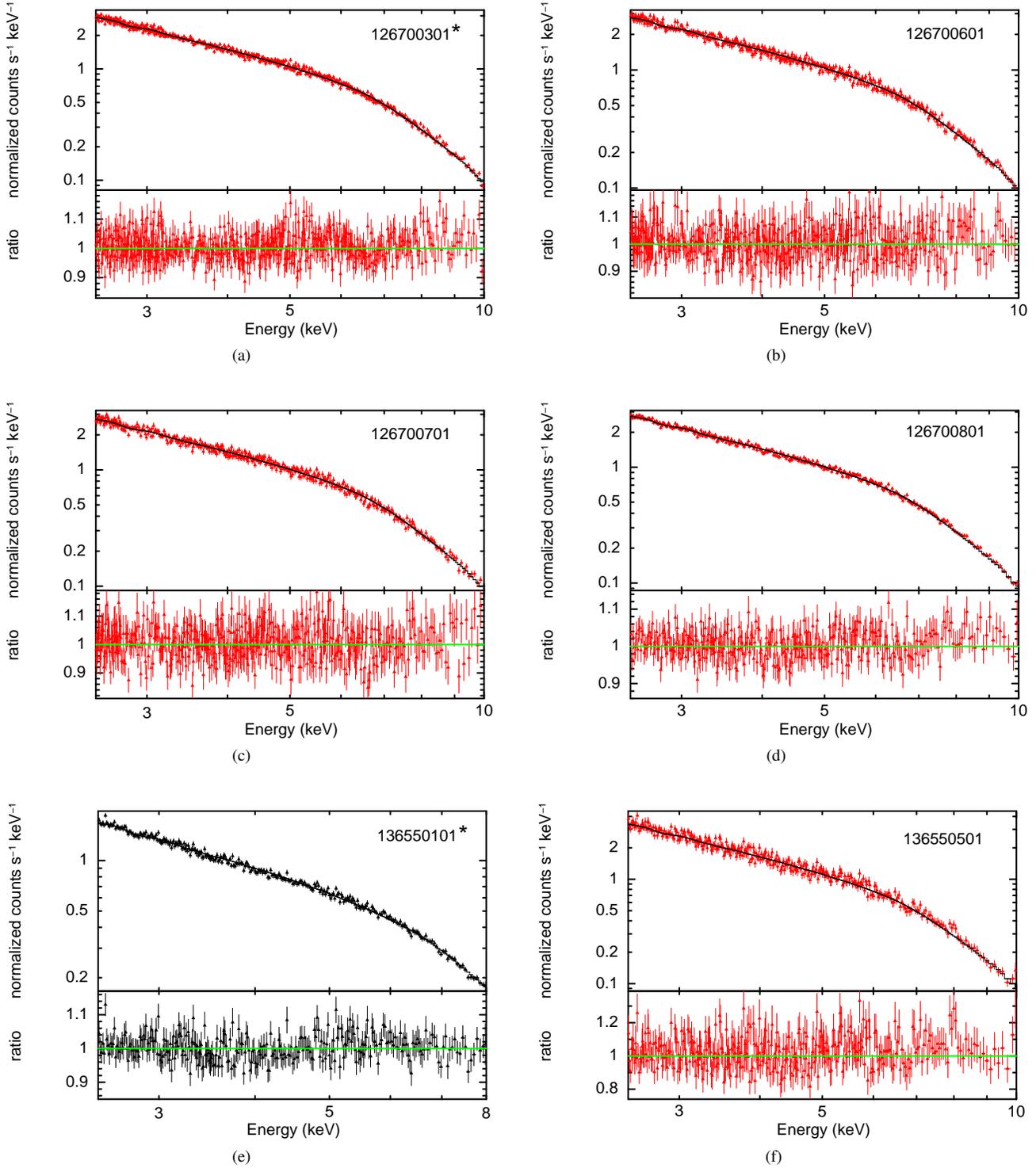

  \centering
  \subfloat[][]{\includegraphics[width=0.33\textwidth,angle=-90]{fig1a.eps}}\quad
  \subfloat[][]{\includegraphics[width=0.33\textwidth,angle=-90]{fig1b.eps}}\\
  \subfloat[][]{\includegraphics[width=0.33\textwidth,angle=-90]{fig1c.eps}}\quad
  \subfloat[][]{\includegraphics[width=0.33\textwidth,angle=-90]{fig1d.eps}}\\
  \subfloat[][]{\includegraphics[width=0.33\textwidth,angle=-90]{fig1e.eps}}\quad
  \subfloat[][]{\includegraphics[width=0.33\textwidth,angle=-90]{fig1f.eps}}\\
  
\caption{The 2.5--10 keV and 2.5--8.0 keV X-ray spectra of 3C 273 as observed by the EPIC-pn camera on board {\it XMM-Newton} 
from 2000--2015 are plotted in red and black, respectively. Observation IDs followed by asterisks show pile up and/or the presence of
a broad iron line component.  The fits are to absorbed power-laws using a Galactic neutral hydrogen column density fixed at 
$N_{H}$ = 1.8 $\times$ 10$^{20}$ cm$^{-2}$, with the lower panels showing the data-to-model ratios. The relevant parameters are 
given in Table 3.}
\end{figure*}

\begin{figure*}
  \centering
  \ContinuedFloat
  \subfloat[][]{\includegraphics[width=0.33\textwidth,angle=-90]{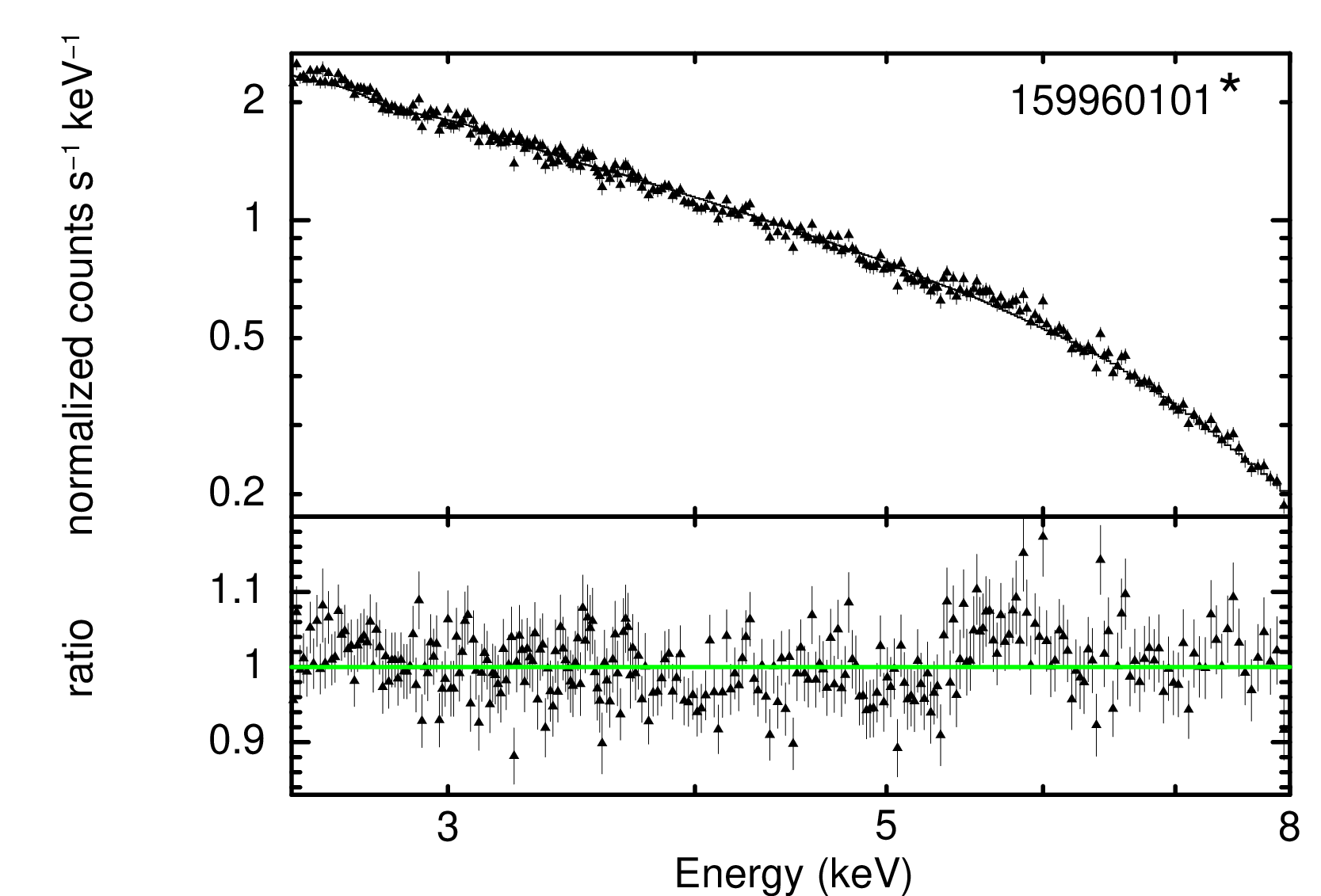}}\quad
  \subfloat[][]{\includegraphics[width=0.33\textwidth,angle=-90]{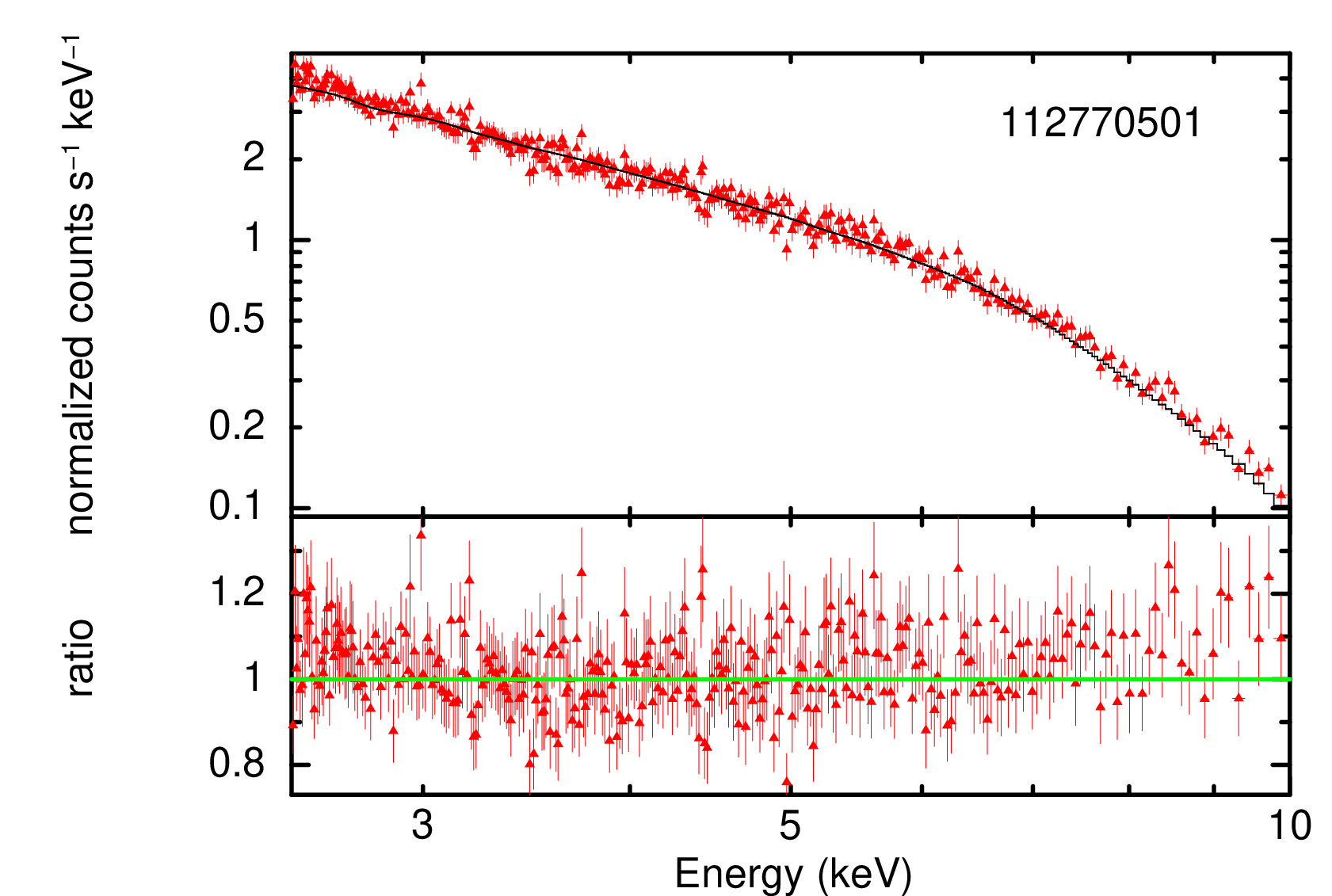}}\\
  \subfloat[][]{\includegraphics[width=0.33\textwidth,angle=-90]{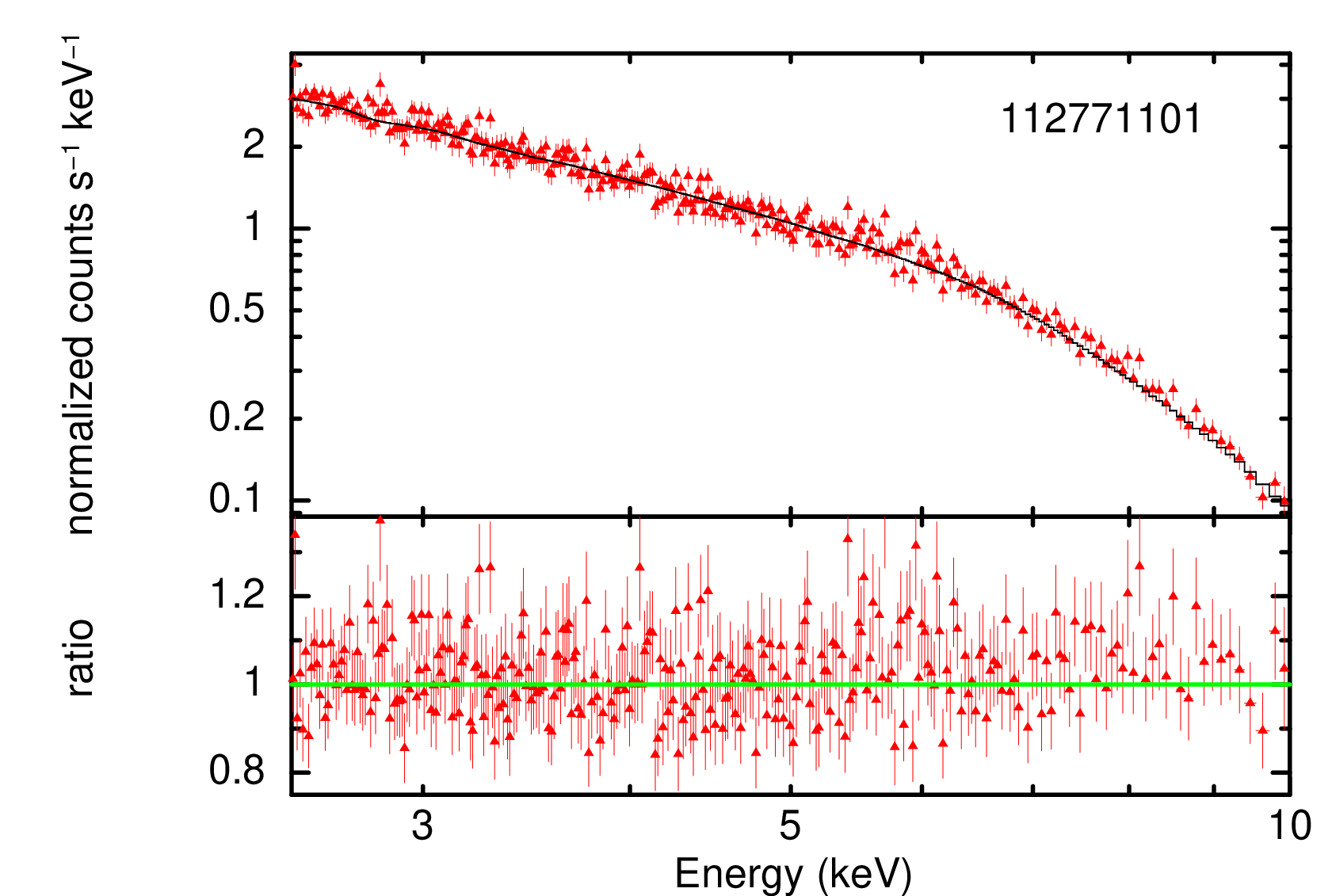}}\quad
  \subfloat[][]{\includegraphics[width=0.33\textwidth,angle=-90]{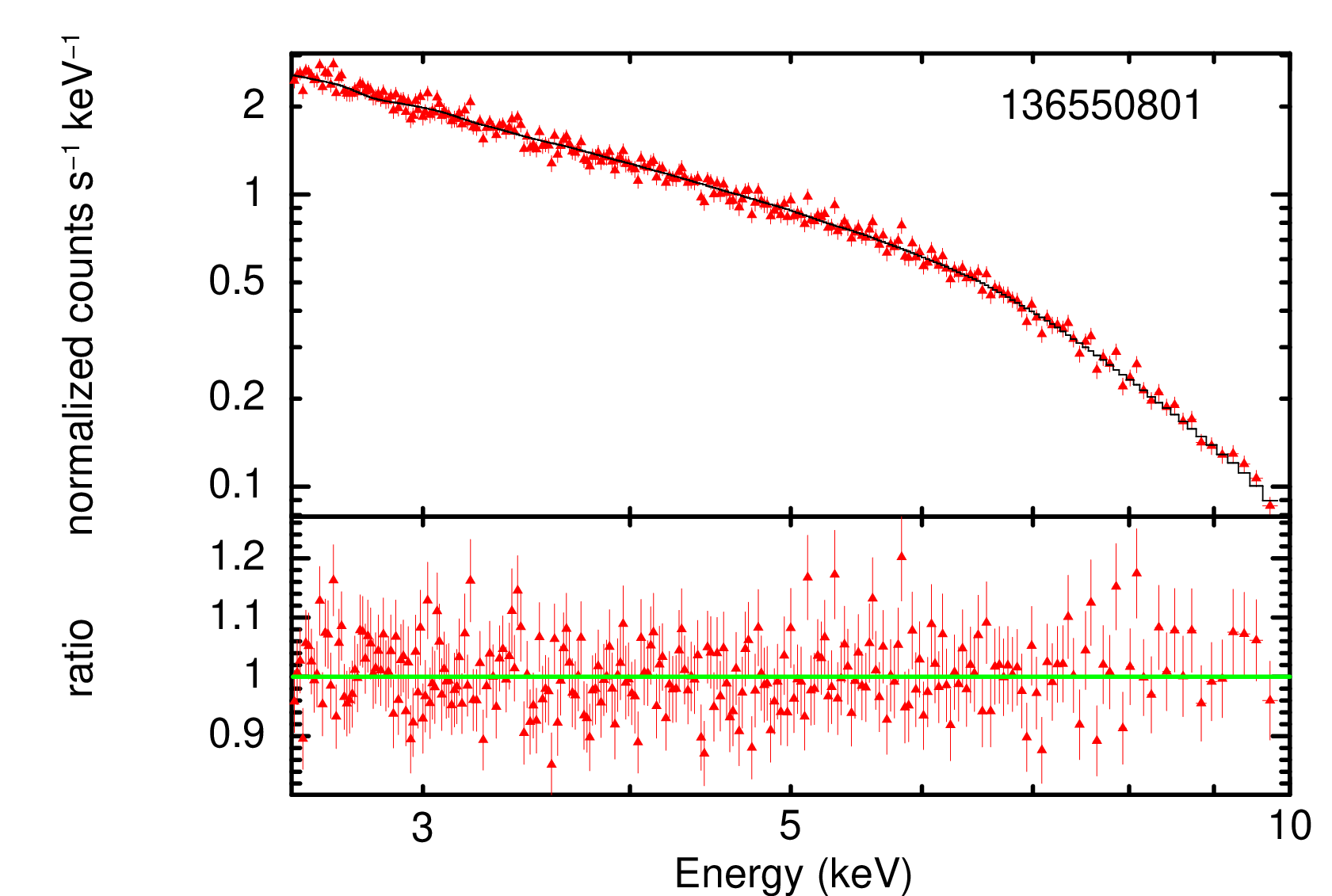}}\\
  \subfloat[][]{\includegraphics[width=0.33\textwidth,angle=-90]{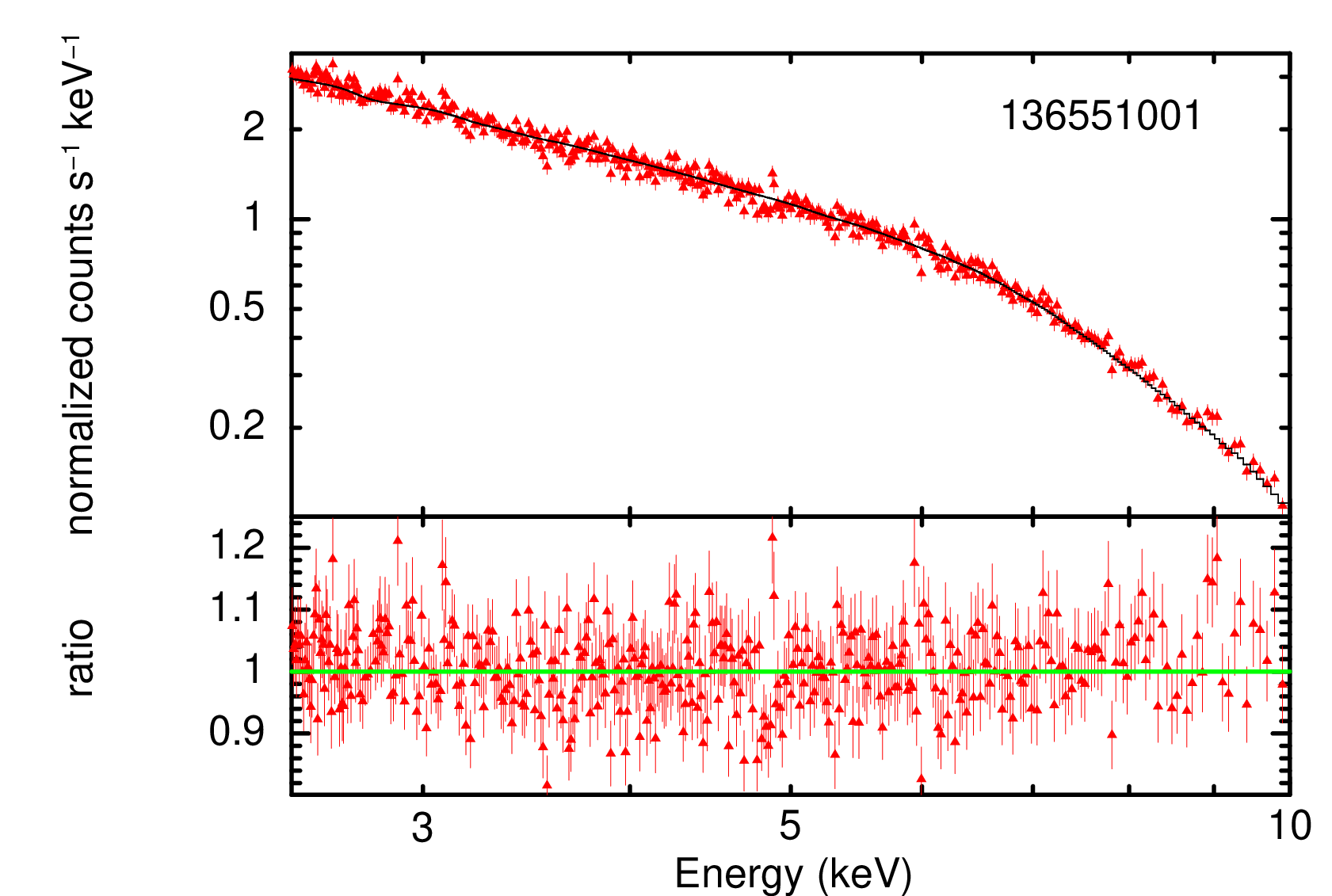}}\quad
  \subfloat[][]{\includegraphics[width=0.33\textwidth,angle=-90]{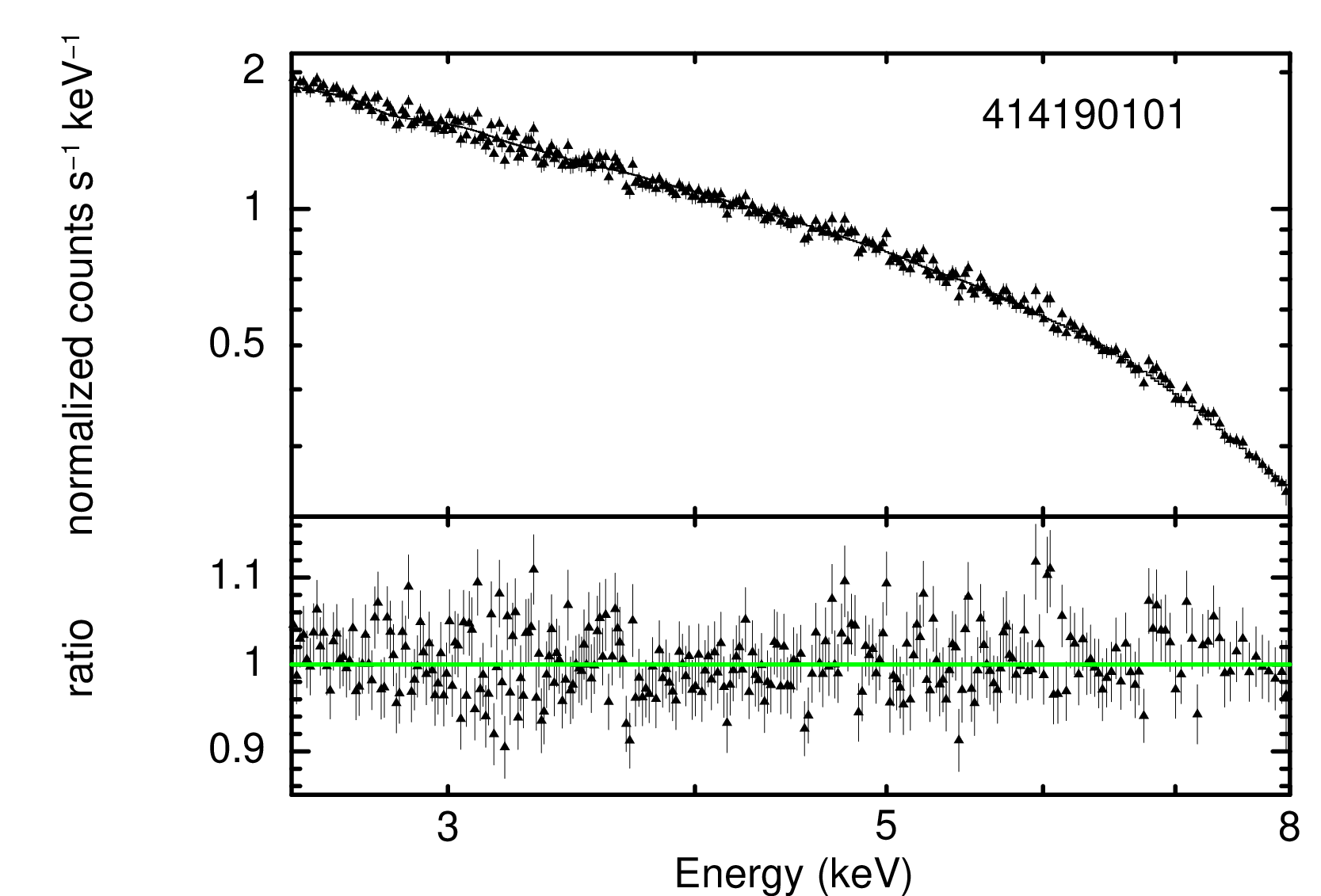}}\\
\caption{continued}
\end{figure*}
 
\begin{figure*}
  \centering
  \ContinuedFloat
  \subfloat[][]{\includegraphics[width=0.33\textwidth,angle=-90]{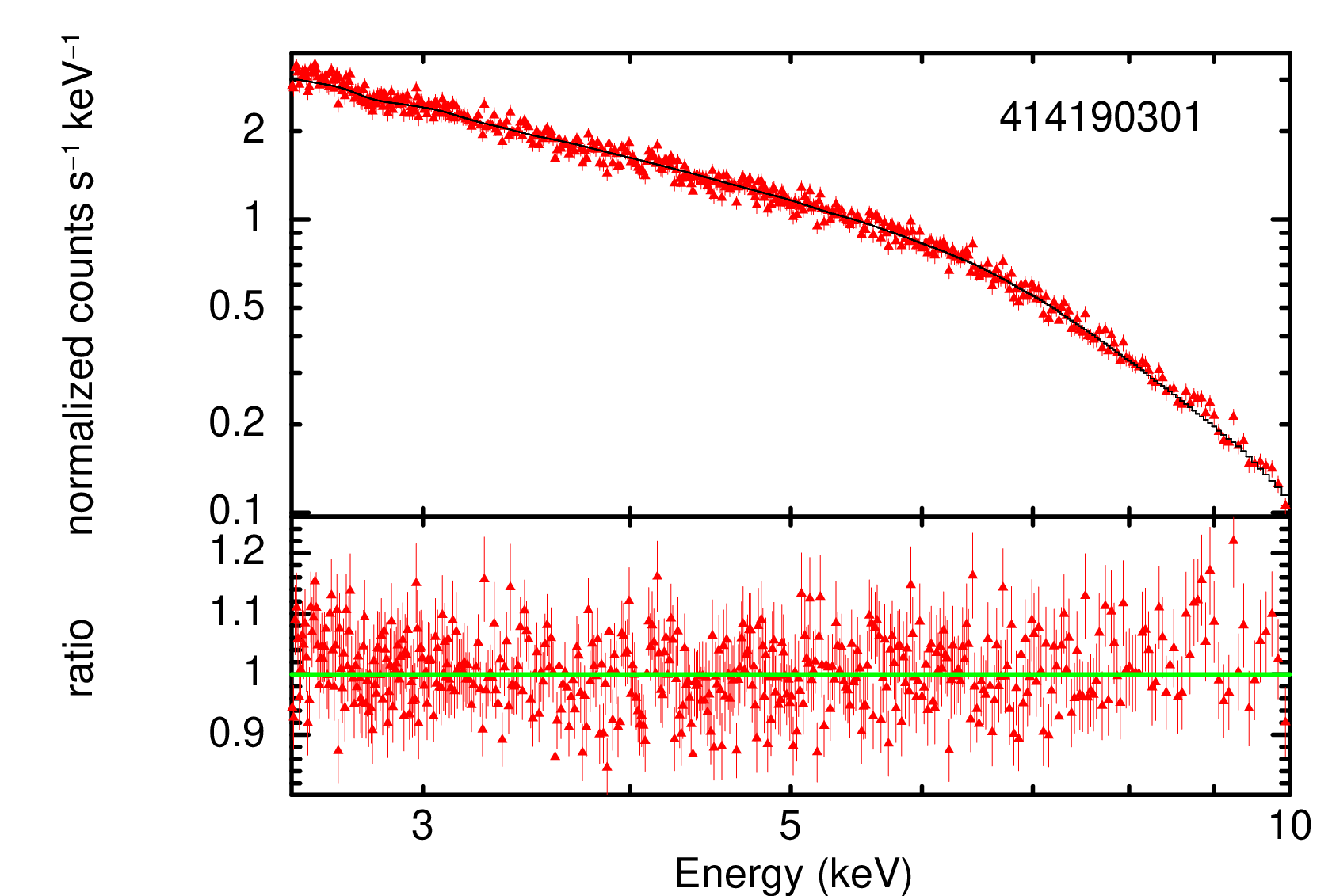}}\quad
  \subfloat[][]{\includegraphics[width=0.33\textwidth,angle=-90]{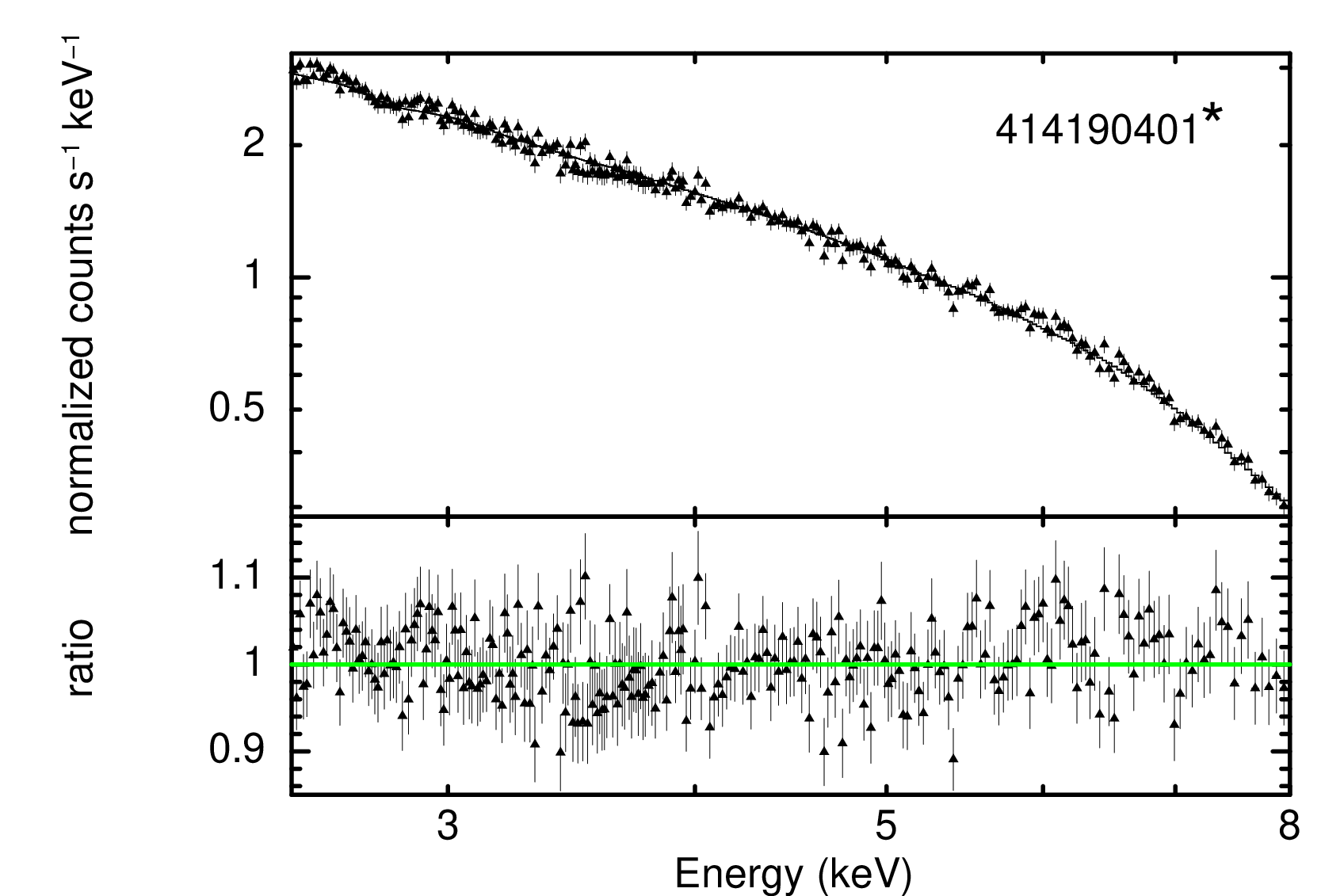}}\\
  \subfloat[][]{\includegraphics[width=0.33\textwidth,angle=-90]{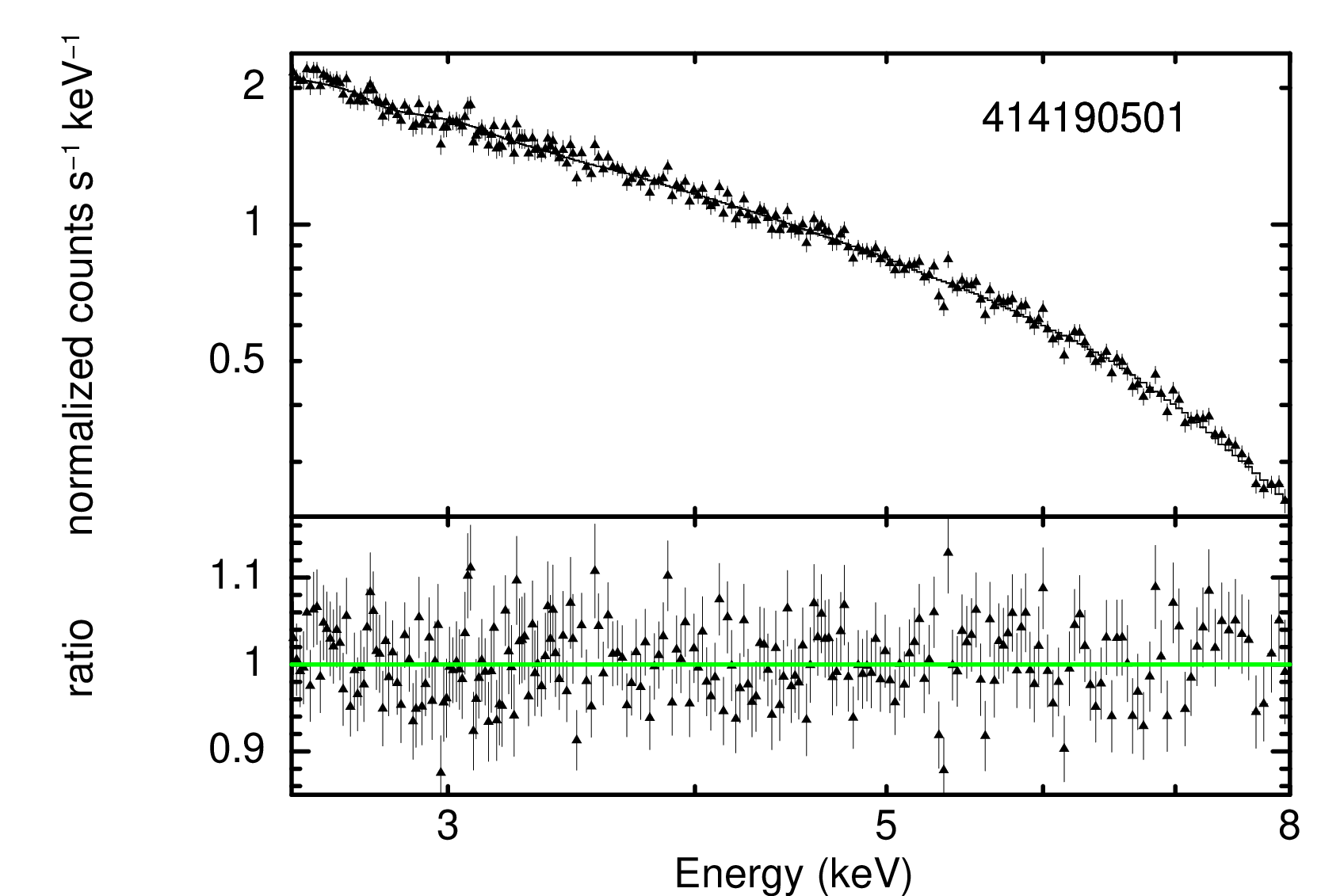}}\quad
  \subfloat[][]{\includegraphics[width=0.33\textwidth,angle=-90]{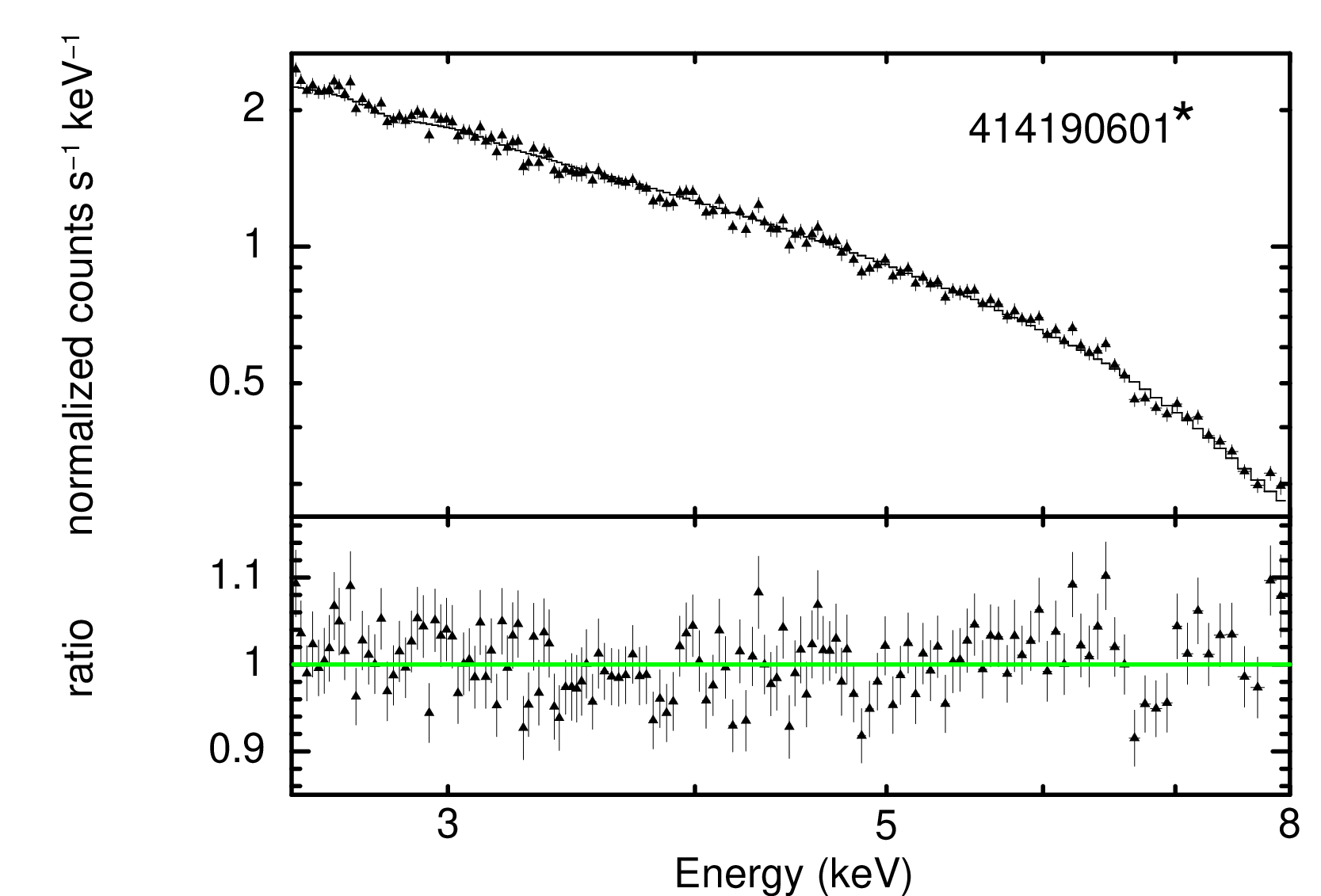}}\\
  \subfloat[][]{\includegraphics[width=0.33\textwidth,angle=-90]{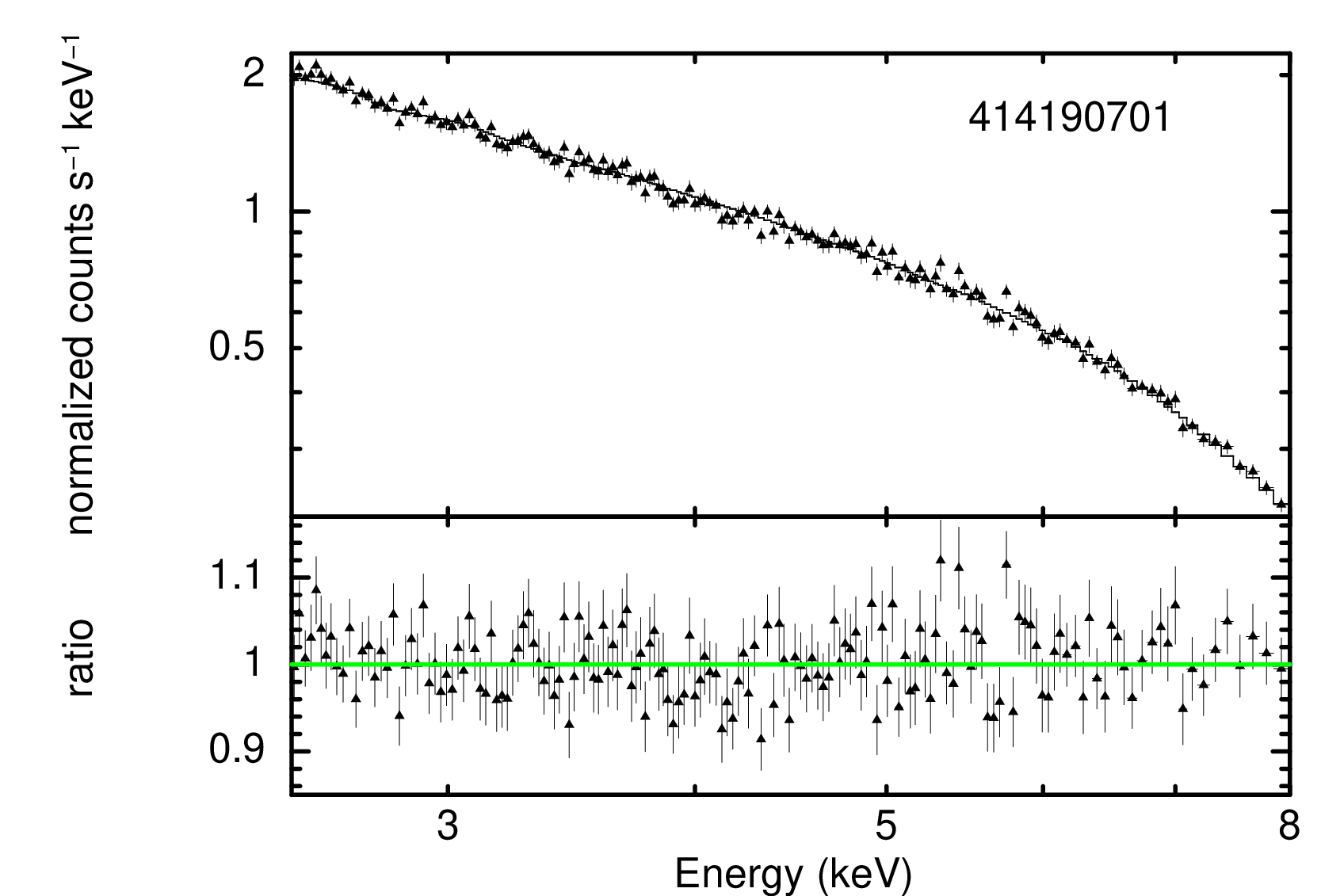}}\quad
  \subfloat[][]{\includegraphics[width=0.33\textwidth,angle=-90]{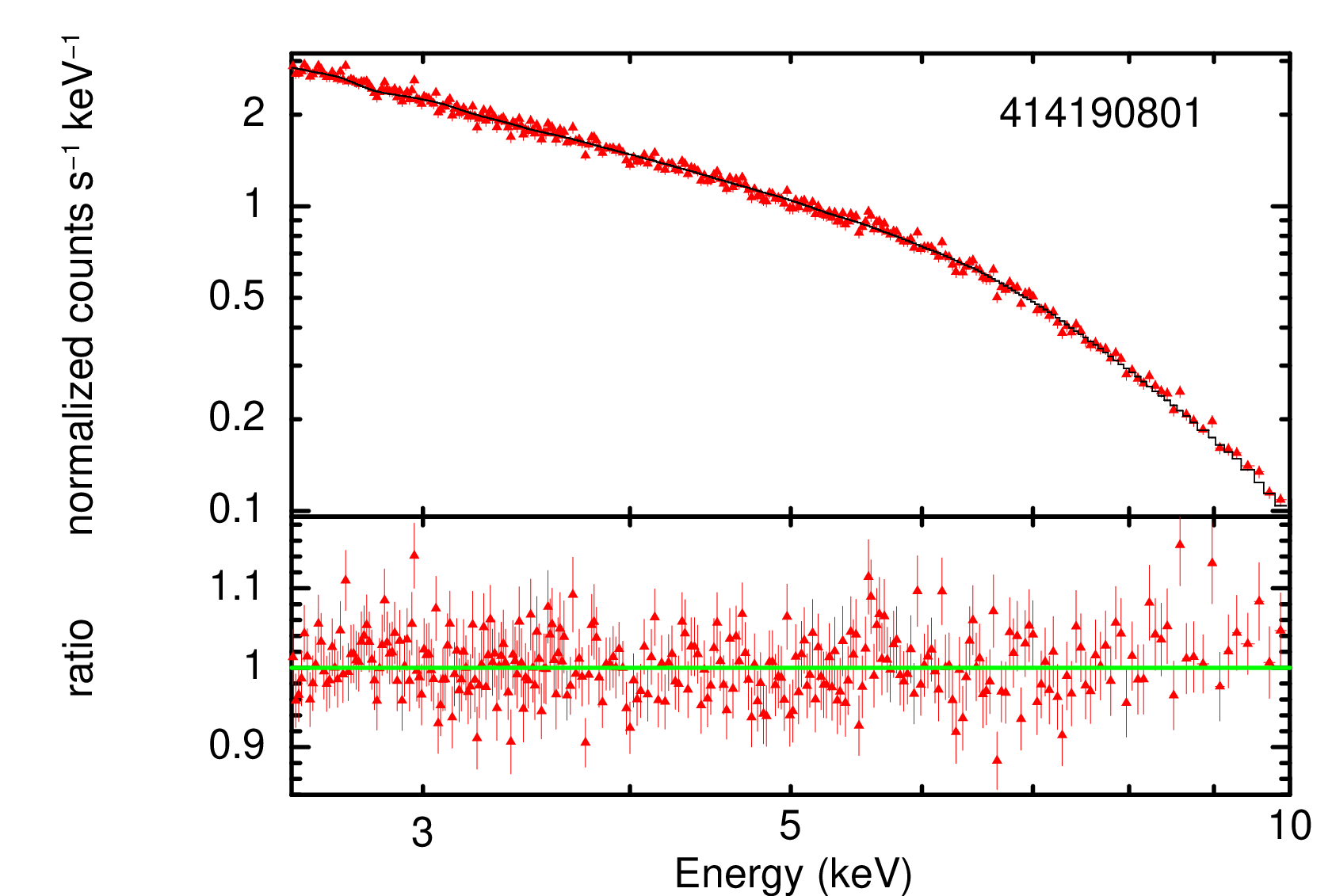}}\\
\caption{continued}
\end{figure*}

\begin{figure*}
  \centering
  \ContinuedFloat
\subfloat[][]{\includegraphics[width=0.33\textwidth,angle=-90]{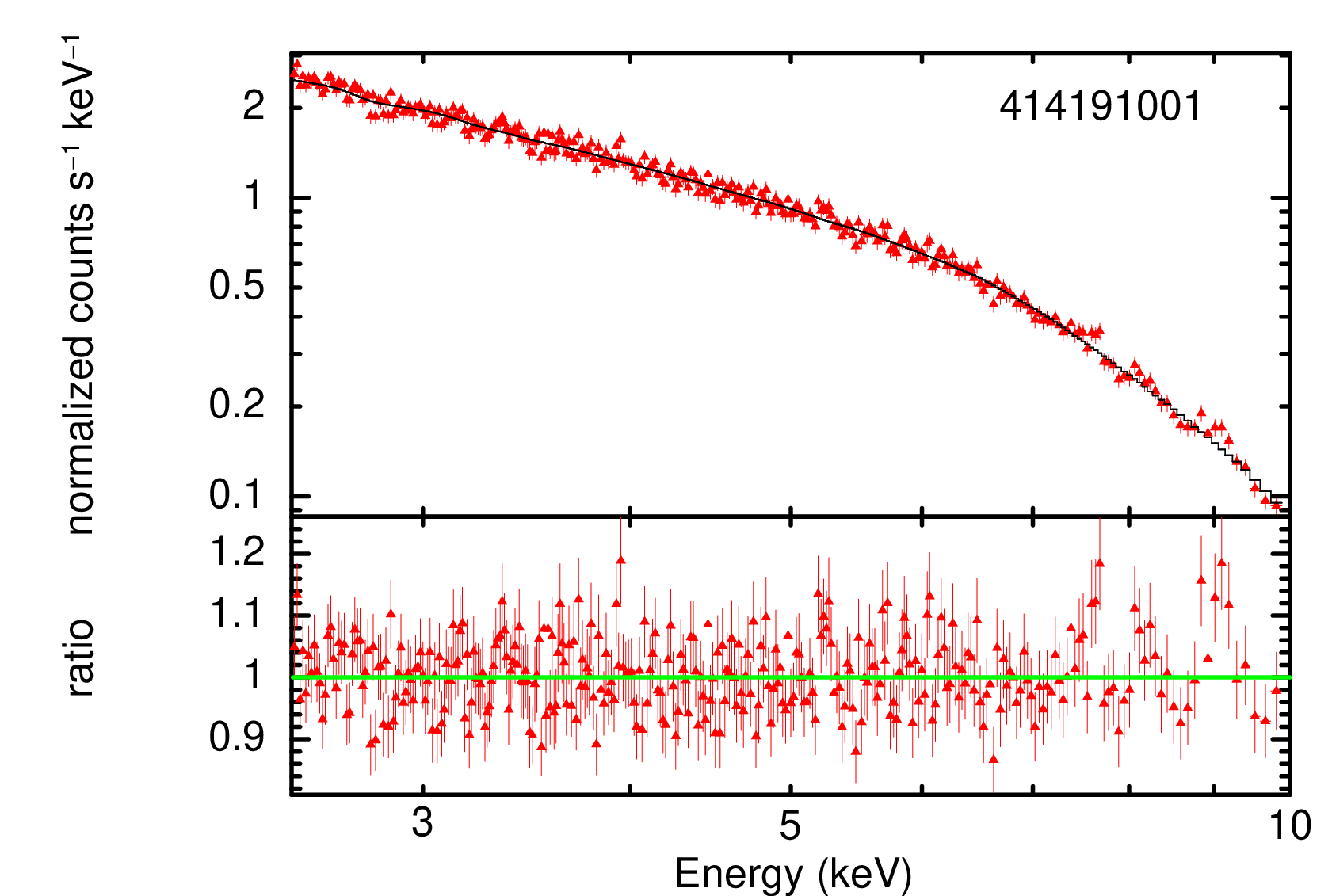}}\quad
\subfloat[][]{\includegraphics[width=0.33\textwidth,angle=-90]{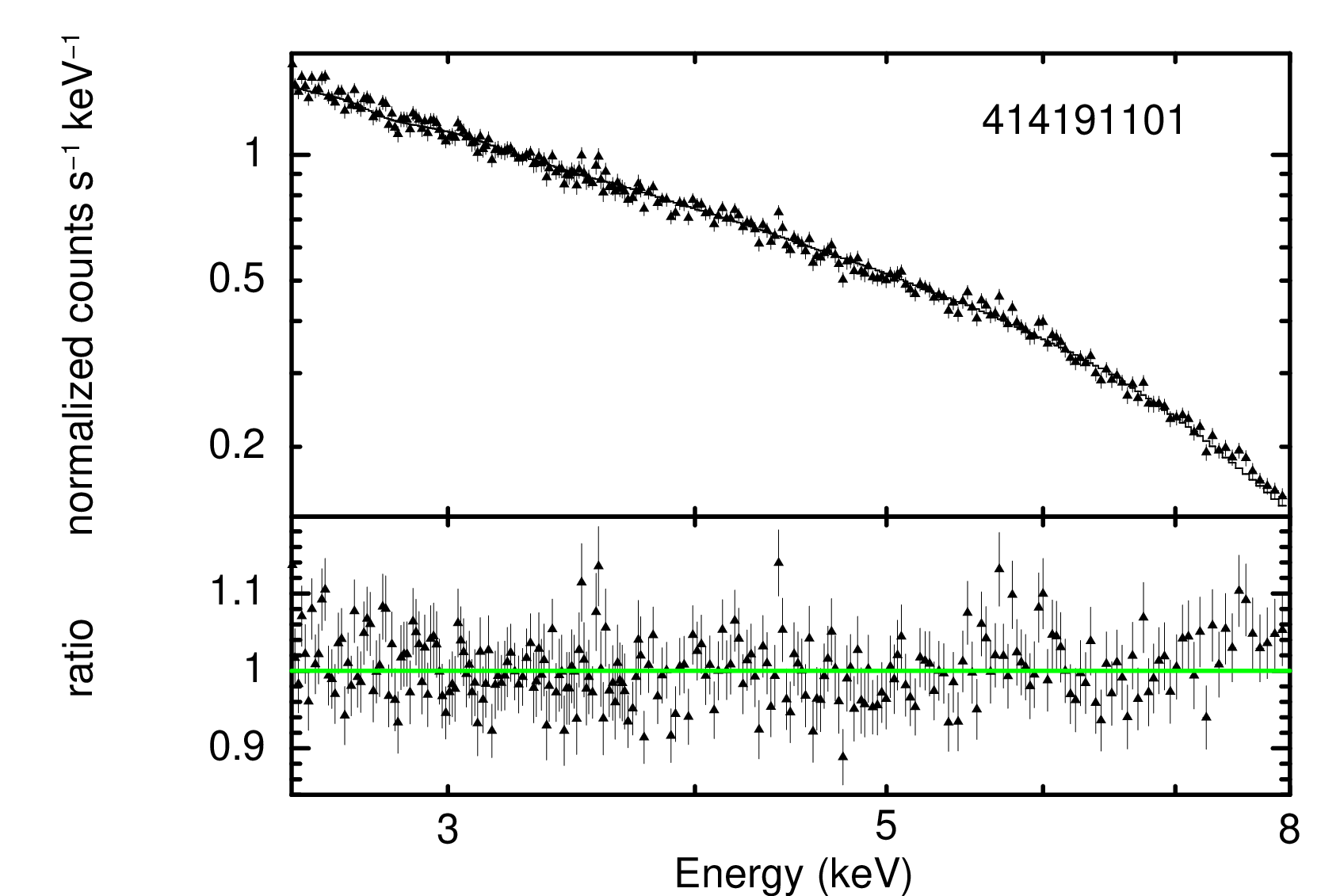}}\\
 \caption{continued}
\end{figure*}

\section {XMM--NEWTON OBSERVATIONS}

\subsection { EPIC-pn Data and X-ray Spectral Analysis}

3C 273 is one of the  sources used to calibrate  {\it XMM-Newton} and has been frequently observed by the satellite from the beginning of the observatory to this date. Hence, {\it XMM-Newton}'s data archive is one of the richest X-ray data sets available for this source. 
Here, we have undertaken the spectral study of 15 years of observations from June 2000 to July 2015, listed in Table 1, for which the photometry was already discussed in Kalita et al.\ (2015). 

The Observational Data Files are collected from the online {\it XMM-Newton} Science Archive. For data reduction, Science Analysis Software (SAS) version-14.0.0 is used. Detailed information about the data and techniques used to reduce the data are discussed\ in Kalita et al.\ (2015), so here we discuss only the methods used for spectral product generation.  Due to sensitivity limits we considered only EPIC-pn observations taken in imaging mode. Lengths of the data trains are required  to be $\geq$ 8 ks, so as to produce statistically significant results.  We only include on-axis observations in this study. This thus excludes the observations made during 2013 which were taken at an offset angle of 8.38$^{\prime}$.  Moreover, these 2013 observations are highly affected by proton flaring.  This constraint leaves us with total number of 20 observations for our study. These observations give a signal-to-noise ratio value of $\sim$146 on average, with lower and upper limits of 62 and 242, respectively.  All these on-axis observations  are  taken with the THIN and MEDIUM filters. 
 
For the source and background spectra,  circular regions of 40 arcsec, centered on the source area and on a source-free region are selected, respectively, using the {\it EVSELECT} task. A spectral binsize of 5 is considered with both single and double events in the energy range 0.3--10 keV.  Seven observations out of 20 are weakly/moderately affected by pile-up problems as reported in Kalita et al.\ (2015) and noted here in Table 1. Pile-up can be minimized to some extent by considering only single pixel events for the spectra, but this method reduces the sensitivity of the data and causes a loss of events. Therefore we have considered both single and double pixel events for spectra generation. The pile-up correction is tackled by removing the central portion of source region where depression of photon counts is maximum. The source and background spectra were extracted from an annulus region for both following the method explained in Molendi $\&$ Sembay (2003). The {\it SAS} task {\it EPATPLOT} is used to check for possible pile-up in the observations. For this, the pattern statistics were first checked for a circular region of 40$^{\prime\prime}$ radius centered on the source. If pile-up is detected, an annular region of inner radius of 6$^{\prime\prime}$ and outer radius of 40$^{\prime\prime}$ was considered first to reduce pile-up and then the pattern plot was rechecked. Depending on the degree of pile-up, the radius of the excluded inner area varies. We gradually increase the inner source radius till the pile up is removed which is determined through a pattern plot. Through this process the inner radius was varied from 6$^{\prime\prime}$--10$^{\prime\prime}$, but the outer radius was fixed at 40$^{\prime\prime}$ for all the pile-up affected observations. The redistribution matrix is then generated using the SAS task {\it rmfgen}. The final background-subtracted spectra were extracted using the SAS task {\it EVSELECT}. The resulting spectra were rebinned with 250 counts per channel using the task {\it grppha} with a oversample upper limit of 3 for good intrinsic energy resolution. 


There are some negative consequences that arise from removing the central source region from the piled-up observations. Pile-up corrections may cause loss of spectral flux and distortion of the spectra at their high energy ends. Thus a high energy break appears near or above 8 keV in those pile-up corrected spectra. To correct for the flux loss and energy distortion caused by the pile-up of photons we use the task {\it rmfgen} which can only be applied for EPIC-pn observations taken in imaging mode. This is done by generating a response file (rmf) from the raw events file which includes the pile-up correction.  Later this rmf was used in spectral analysis. Even after applying this correction, the break remains to some degree and can be seen from the residual plots in Figs.\ 1 and 2. Even using only single event spectra we cannot completely eliminate the pile-up problem and both the flux loss problem and the high energy break remain regardless. So, without compromising the sensitivity, we chose to take both single and double events for spectra generation.

\subsection{Optical Monitor (OM) Data}

Apart from its high sensitivity X-ray telescopes, {\it XMM-Newton} has an additional observing facility, a 30 cm aperture Ritchey-Chr{\'e}tien optical/UV telescope with focal length of 3.8 m with time resolution of 0.5 second. The main purposes of this OM are to provide accuracy in tracking objects and to observe the target simultaneously in optical/UV bands with the X-ray telescopes.  The OM provides coverage between 170 to 650 nm over a 17 arc minute square region of the X-ray field of view. The detector contains a set of broad band filters, 3 in UV bands and 3 in optical bands, with effective wavelengths of 212, 231, and 291 nm and 344, 450, and 543 nm, respectively. (For more details see Mason et al.\ 2001). Out of a total of 20 observations we use in this work, 17 contain simultaneous optical data (see Table 1). Optical/UV data were processed with the {\it omichain} pipeline, as instructed in the SAS Analysis thread\footnote{http://www.cosmos.esa.int/web/xmm-newton/sas-threads}. The count rate for respective filters are collected from the final source list result files. If we assume that the X-ray power-law emission is an extrapolation of low energy photons (optical/UV) via Comptonization processes then the high energy power-law cutoff will represent the maximum energy range of the emitted X-ray photons. For study of correlations between bands we consider the highest frequency UV filter, UVW2 (6.2 eV), through which simultaneous X-ray/UV observations were taken 14 times by {\it XMM-Newton}. Then the count rates are  converted to flux based on white dwarf standard star observations (using method 1 as instructed on the SAS watchout page\footnote{http://www.cosmos.esa.int/web/xmm-newton/sas-watchout-uvflux}). The resulting fluxes for this filter are provided in Table 2.

\begin{table}

{\bf Table 2. Optical Monitor UVW2 filter log} \\
\small
\begin{tabular}{lccc} \hline \hline
Date of obs.&  Obs. ID &  $F_{\rm UVW2}^{1}$         &  $F_{err}^{2}$    \\
(dd.mm.yyyy)&          &(erg cm$^{-2}$ s$^{-1}$)&(erg cm$^{-2}$ s$^{-1}$) \\\hline \hline
 
 13.06.2000 &126700301 &  3.3387364e-10 & 1.2408036e-13\\     
 15.06.2000 &126700601 &  --- $^{\star}$       		& ---     \\      
 15.06.2000 &126700701 &  3.3251290e-10 & 1.3254516e-13\\      
 17.06.2000 &126700801 &  3.3222661e-10 & 1.2548698e-13\\      
 13.06.2001 &136550101 &  4.0486519e-10 & 1.4448513e-13\\      
 05.01.2003 &136550501 &  --- 		       & ---     \\
 07.07.2003 &159960101 &  3.8417051e-10 & 1.7985725e-13\\      
 08.07.2003 &112770501 &  --- 		       & ---     \\
 14.12.2003 &112771101 &  --- 		       & ---     \\
 30.06.2004 &136550801 &  3.1228535e-10 & 2.1105259e-13\\      
 10.07.2005 &136551001 &  3.6505119e-10 & 2.9180965e-13\\      
 12.01.2007 &414190101 &  3.2662444e-10 & 1.6241971e-13\\
 25.06.2007 &414190301 &  2.8009823e-10 & 1.9517117e-13\\     
 08.12.2007 &414190401 &  3.9972460e-10 & 2.9177285e-13\\    
 09.12.2008 &414190501 &  3.1919197e-10 & 2.6378211e-13\\     
 20.12.2009 &414190601 &  --- 		       & ---     \\      
 10.12.2010 &414190701 &  3.3076551e-10 & 2.4767070e-13\\     
 12.12.2011 &414190801 &  3.4609663e-10 & 2.4635401e-13\\     
 16.07.2012 &414191001 &  --- 		     & ---      \\
 13.07.2015 &414191101 &  2.2965041e-10 & 1.0593299e-12\\\hline
\end{tabular}     \\
$^{1}$  UV flux of UVW2 filter, $^{2}$ Corresponding error in flux \\
$^{\star}$ Note:  ---  indicates that either there were 
no OM observations or the \\UVW2 exposure is absent. \\
\end{table}

\section {SPECTRAL FITTING WITH XSPEC}

For the spectral analysis we have used the X-ray spectral fitting package {\it XSPEC} version 12.8.2 provided by High Energy Astrophysics Science Archival Research Center (HEASARC), NASA/GSFC. The goodness of fit for the spectral fitting process is determined by  the reduced $\chi^{2}$ value, since the number of data points in each spectra are large enough to satisfy the condition for use of $\chi^{2}$ statistics. We have performed spectral fitting for all the spectra in the energy range 2.5--10 keV, since our aim is to study the harder continuum of the source. The lower limit of 2.5 keV is employed to avoid any contribution from the complicated soft excess emission which is influenced by galactic and other soft X-ray emissions found in the object. In all our fitting analyses we have considered X-ray absorption due to Galactic neutral hydrogen towards the observer. Unless stated otherwise, all the errors quoted are $2\sigma$ (90$\%$). We report only those fits with reduced $\chi_{r}^{2}$ less than 1.5. For any added spectral component, the improvement of the fit is determined by both the change in the reduced $\chi_{r}^{2}$ value,  $\Delta \chi_{r}^{2}$, and the F-test value. We use the  in-built tool in the XSPEC package to compute $\Delta \chi_{r}^{2}$ which gives the improvement of the spectral fitting after adding an new model component to the first model used for fitting the spectra. It calculates the F-statistic and its probability from the two $\chi^{2}$ values computed from those two different fits.

\subsection{X-ray Spectra: single power-law model} 

We first fit the spectra with a simple power-law absorbed by the Galactic hydrogen column along the line of sight to 3C 273. For all the fits we fixed the neutral  H column density value at  $N_{H}$ $= 1.8 \times 10^{20}$ cm$^{-2}$ (Dickey $\&$ Lockman 1990)
and from the website\footnote{https://heasarc.gsfc.nasa.gov/cgi-bin/Tools/w3nh/w3nh.pl} (nH Column Density-HEASARC-NASA). 
The model
specifications are provided by Wilms et al.\ (2000). Out of 20 spectra, 13 are nicely fit by power-laws alone, with statistically excellent reduced $\chi^{2}$ values. Also, the deviation of the data/model ratio from unity lies within a 20$\%$ error limit for those fits.  Figure 1 contains all of the spectra fitted with an absorbed power-law. The power-law spectral indices, $\Gamma$, have a minimum slope of 1.35 in January 2007 and a maximum slope 1.79 in July 2003 and thus show a range of $\Delta\Gamma =$ 0.44. For those observations affected by the pile-up problem, a break appears in the spectra at or above 8 keV, which can be seen most easily from the data/model ratio plots. These ``breaks''  occur due to the exclusion of the central source region to remove the pile-up effect from the light curves. 
Though we have taken care of the flux loss and energy distortion caused by the pile-up correction by regenerating the redistribution matrix as mentioned earlier, it seems the  distortion at the high energy end of the spectrum cannot be overcome fully.  Despite the breaks near the tails of these spectra, power-laws still give statistically acceptable fits, albeit typically with slightly higher $\chi_{r}^{2}$ values as compared to the rest (see Table 3) with totally reasonable parameters. The X-ray fluxes are calculated within the energy range 2.5--10 keV using a simple power-law model.

\begin{figure*}
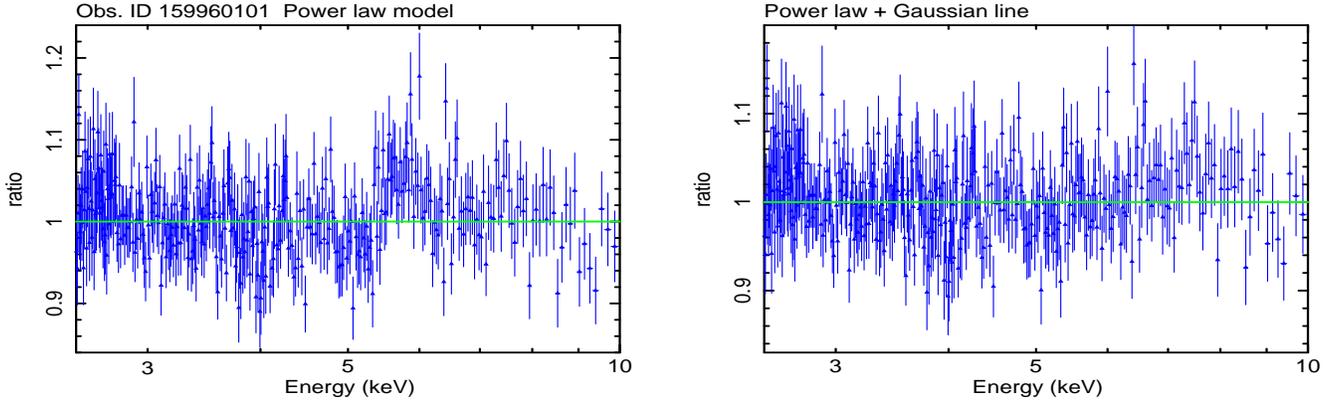
      
\centering
\mbox{\subfloat{\includegraphics[width=0.3\textwidth,height=0.37\textheight,angle=-90]{fig2a.eps}}\quad
\subfloat{\includegraphics[width=0.3\textwidth,height=0.37\textheight,angle=-90]{fig2b.eps} }}
\caption{Data to model ratios of the Observation 0159960101. Left: A simple power-law fit to the spectrum with galactic absorption. At near 6 keV the residual shows a bump indicating the presence of an iron line. Right: The bump disappears from the residuals after adding a Gaussian line to the power-law spectra; addition of a Gaussian line improves the fit by changing $\chi_{r}^{2}$ value from 1.13 to 1.05 ($\Delta$ $\chi^{2}$ = 48/3 degrees of freedom).}
\vspace*{0.7cm}
\end{figure*}

\begin{table*}
{\bf Table 3. The parameters of 2.5--10 keV X-ray continuum extracted by fitting simple power-laws  }\\ 
{\bf with fixed galactic absorption, N$_{H}$ = 1.8 $\times$ 10$^{20}$ cm$^{-2}$}\\
\small
\begin{tabular}{lccclccc} \hline \hline
Date of obs.&  Obs. ID &    $\Gamma^{a}$ &      N$^{b}$         &$\chi^{2}$/DOF&   $\chi_{r}^{2}$&Null- &  ~~~~$F^{c}$ \\
(dd.mm.yyyy)&          &                 &     ($\times$ 10$^{-02}$)     &              &          &-hypothesis & (erg cm$^{-2}$ s$^{-1}$)\\\hline

13.06.2000 &126700301$^\star$ &  1.64$_{-0.01}^{+0.01}$  & 1.75$_{-0.02}^{+0.02}$ &482.81/469 & 1.03  &3.20e-01  &7.06e-11 \\
\\
15.06.2000 &126700601  &  1.59$_{-0.01}^{+0.01}$  & 1.61$_{-0.03}^{+0.03}$ &451.04/425 & 1.06  &1.84e-01  &7.03e-11\\
\\
15.06.2000 &126700701  &  1.60$_{-0.01}^{+0.01}$  & 1.58$_{-0.03}^{+0.03}$ &444.78/421 & 1.06  &2.04e-01  &6.83e-11\\
\\
17.06.2000 &126700801  &  1.61$_{-0.01}^{+0.01}$  & 1.63$_{-0.02}^{+0.02}$ &696.55/693 & 1.01  &2.26e-02  &6.81e-11\\
\\
13.06.2001 &136550101$^\star$  &  1.54$_{-0.01}^{+0.01}$  & 2.10$_{-0.03}^{+0.03}$ &709.14/693 & 1.02  &2.26e-02  &9.84e-11\\
\\
05.01.2003 &136550501  &  1.77$_{-0.02}^{+0.02}$  & 2.29$_{-0.07}^{+0.07}$ &932.05/917 & 1.02  &3.57e-01  &7.45e-11\\
\\
07.07.2003 &159960101$^\star$  &  1.72$_{-0.01}^{+0.01}$  & 2.54$_{-0.05}^{+0.05}$ &677.44/600 & 1.13  &1.52e-02  &8.98e-11\\
\\
08.07.2003 &112770501  &  1.79$_{-0.02}^{+0.02}$  & 2.62$_{-0.09}^{+0.09}$ &150.75/159 & 0.95  &6.67e-01  &8.23e-11\\
\\           
14.12.2003 &112771101  &  1.66$_{-0.02}^{+0.02}$  & 1.85$_{-0.07}^{+0.07}$ &143.59/145 & 0.99  &5.18e-01  &7.14e-11\\
\\
30.06.2004 &136550801  &  1.70$_{-0.02}^{+0.02}$  & 1.65$_{-0.04}^{+0.04}$ &262.88/267 & 0.98  &5.60e-01  &5.89e-11\\
\\
10.07.2005 &136551001  &  1.57$_{-0.01}^{+0.01}$  & 1.69$_{-0.03}^{+0.03}$ &508.42/430 & 1.18  &5.40e-03  &7.57e-11\\
\\
12.01.2007 &414190101  &  1.35$_{-0.01}^{+0.01}$  & 2.21$_{-0.03}^{+0.03}$ &777.89/734 & 1.06  &7.50e-02  &1.44e-10\\
\\
25.06.2007 &414190301  &  1.55$_{-0.01}^{+0.01}$  & 1.69$_{-0.03}^{+0.03}$ &580.01/504 & 1.15  &1.06e-02  &7.84e-11\\
\\
08.12.2007 &414190401$^\star$  &  1.55$_{-0.01}^{+0.01}$  & 4.04$_{-0.07}^{+0.08}$ &517.90/523 & 0.99  &4.44e-01  &1.93e-10\\
\\
09.12.2008 &414190501  &  1.44$_{-0.02}^{+0.02}$      & 2.08$_{-0.05}^{+0.05}$ &523.67/459 & 1.14  &3.41e-01  &1.16e-10\\
\\
20.12.2009 &414190601$^\star$&  1.44$_{-0.02}^{+0.02}$& 2.33$_{-0.05}^{+0.05}$ &421.90/388 & 1.09  &8.97e-03  &1.24e-10\\
\\
10.12.2010 &414190701  &  1.48$_{-0.02}^{+0.02}$      & 1.81$_{-0.05}^{+0.05}$ &355.36/382 & 0.93  &8.70e-02  &9.43e-11\\
\\
12.12.2011 &414190801  &  1.61$_{-0.01}^{+0.01}$      & 1.65$_{-0.03}^{+0.03}$ &593.87/597 & 0.99  &5.28e-01  &7.00e-11\\
\\
16.07.2012 &414191001  &  1.60$_{-0.02}^{+0.02}$      & 1.42$_{-0.03}^{+0.03}$ &333.80/331 & 1.01  &4.47e-01  &6.16e-11\\
\\
13.07.2015 &414191101 &  1.61$_{-0.01}^{+0.01}$      & 1.18$_{-0.01}^{+0.00}$ &533.68/505 & 1.06  &1.82e-01  &4.98e-11\\\hline
\end{tabular}     \\
$^{a}$ Power-law spectral index, $^{b}$ Power-law normalization, $^{c}$ Spectral flux in the range 2.5--10 keV.\\
Note: A very broad line feature is likely present in the observations marked with a $^\star$.\\
\end{table*}

\subsection{ Fe K$\alpha$ Emission Line}

The presence of a fluorescence Fe K$\alpha$ line near its rest energy of 6.4 keV has been detected in 3C 273 from time to time with different X-ray observations. The iron line can be neutral  (Turner et al.\ 1990; Page et al.\ 2004; Grandi $\&$ Palumbo 2004) but at other times it can be ionized (Yaqoob \& Serlemitsos 2000; Kataoka et al.\ 2002).  The detailed properties of this line are governed by the intense gravity and Doppler velocity of the emitting circumnuclear material in the vicinity of the supermassive black hole, but there is no clear understanding of  the origin of this line. In our analysis we have detected a relatively narrow Fe line in the observation 0159960101 on 2003 July 7. The ratio plot of this spectra with an absorbed power-law is shown in Figure 2 and the peak-like deviation in the residual from the constant reference line near 6 keV is clearly visible from the first plot. As mentioned earlier, a high energy break is also visible as result of pile-up. 

To most simply model this emission features we added a Gaussian component to the power-law spectrum and allowed the parameters to vary. This gives a similar spectral slope ($\Gamma =1.73$) as found with a single power-law without adding the Gaussian line. The peak energy of the line lies at $5.79_{-0.07}^{+0.08}$ keV, which corresponds to a line width of $\sigma =0.220$ keV and equivalent width, EW = $47.39_{-16}^{+13}$ eV.  The line energy estimated here indicates that the material emitting the Fe K$\alpha$ line in 3C 273 is neutral or mildly ionized. The width of the Gaussian line found here is similar to those of Seyfert  1 galaxies (Nandra et al.\ 1997).  The line flux is calculated with the {\it cflux} task in {\it XSPEC}, which gives a flux value $\thickapprox 5.44 \times 10^{-13}$ ergs cm$^{-2}$ s$^{-1}$, where the overall spectral flux remains exactly the same as the previous value of $8.98 \times 10^{-11}$ ergs cm$^{-2}$ s$^{-1}$ (see Table 2). The addition of the line improves the fit by $\Delta$ $\chi^{2}$ = 48 at the expense of 3 degrees of freedom and the $\chi_{r}^{2}$ value improves to 1.05 from 1.13.   This improvement is 99.99$\%$ significant (with rejection probability of $2.0072 \times 10^{-9}$) given by the F-test with a F-statistic value of $\thickapprox 15$.
A Gaussian line follows the relation, FWHM/$\sigma_{line}$ $\thickapprox 2.355$, which yields a FWHM $\textgreater$ 20,000 km s$^{-1}$ ($\sim 0.08 c$) for this line. Such a broad line width is unlikely to originate from broad line region clouds or be due to the irradiation of high energy photons on to the cold dusty torus (Nandra 2006) but more probably originates from a Seyfert like component: the inner accretion disc (Reynolds 2001).

Apart from this quite clear observation of an emission line, four other observations apparently show excess emission at energies between 5--7 keV in the observer's frame, but these lines are too weak to be considered as significant detections. However, broad Gaussian line fits to these spectra give modest improvements to the $\chi_{r}^{2}$ values and as well as the residual plots. These observations are marked with a star symbol in Table 3.  Although we tried to fit these putative broad lines with accretion disc models for both Schwarzschild geometry (Fabian et al.\ 1989) and Kerr geometry (Laor 1991)  the statistics are too poor to allow quoting any results for these fits.

\begin{figure}
\centering
\psfig{figure = 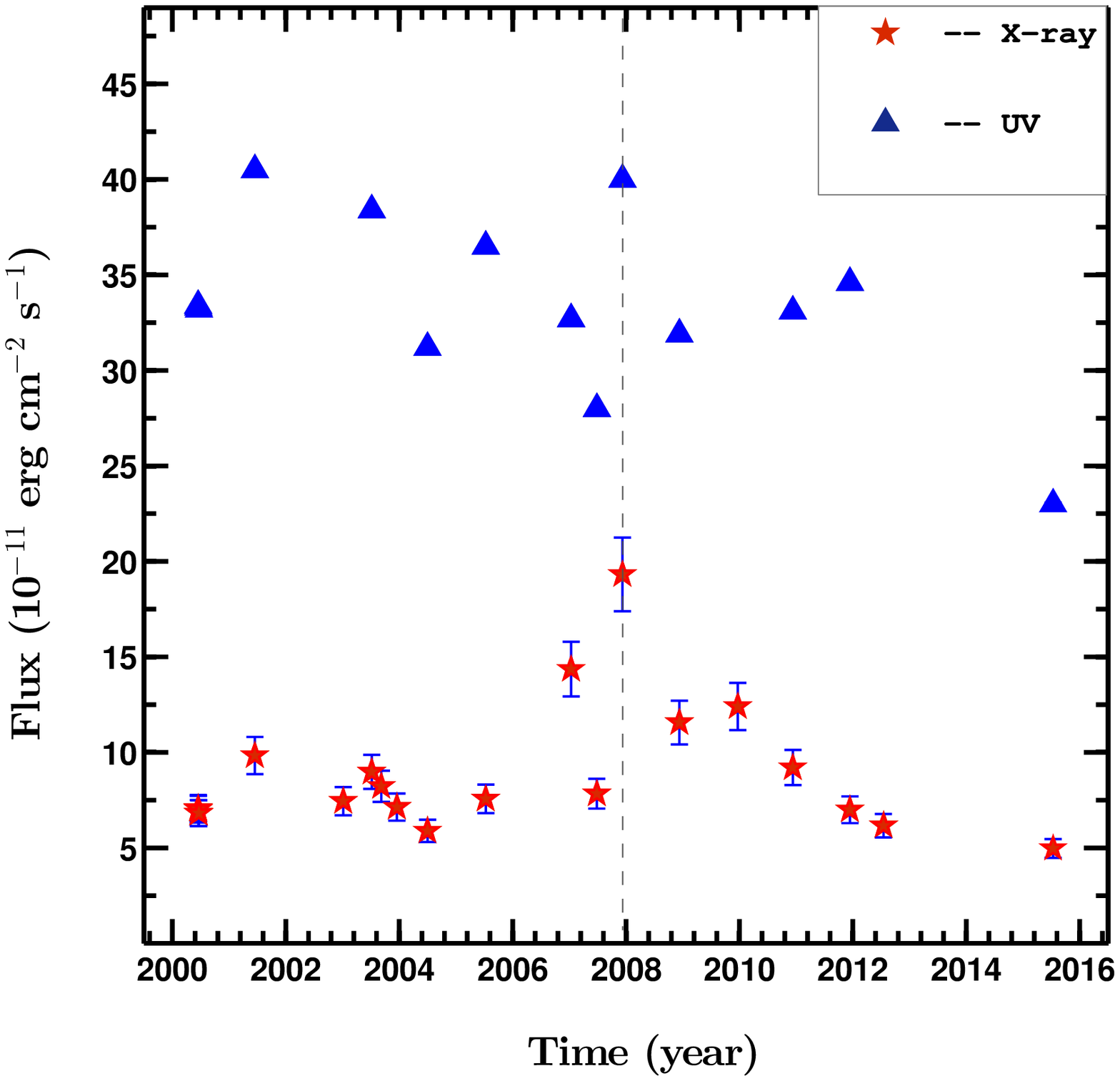 ,scale=0.5}
\caption{Temporal variation of the X-ray and UV fluxes. The vertical dotted line emphasizes the flaring period.}
\label{fig9}
\vspace*{.8cm}
\end{figure}


\begin{figure}
\psfig{figure = 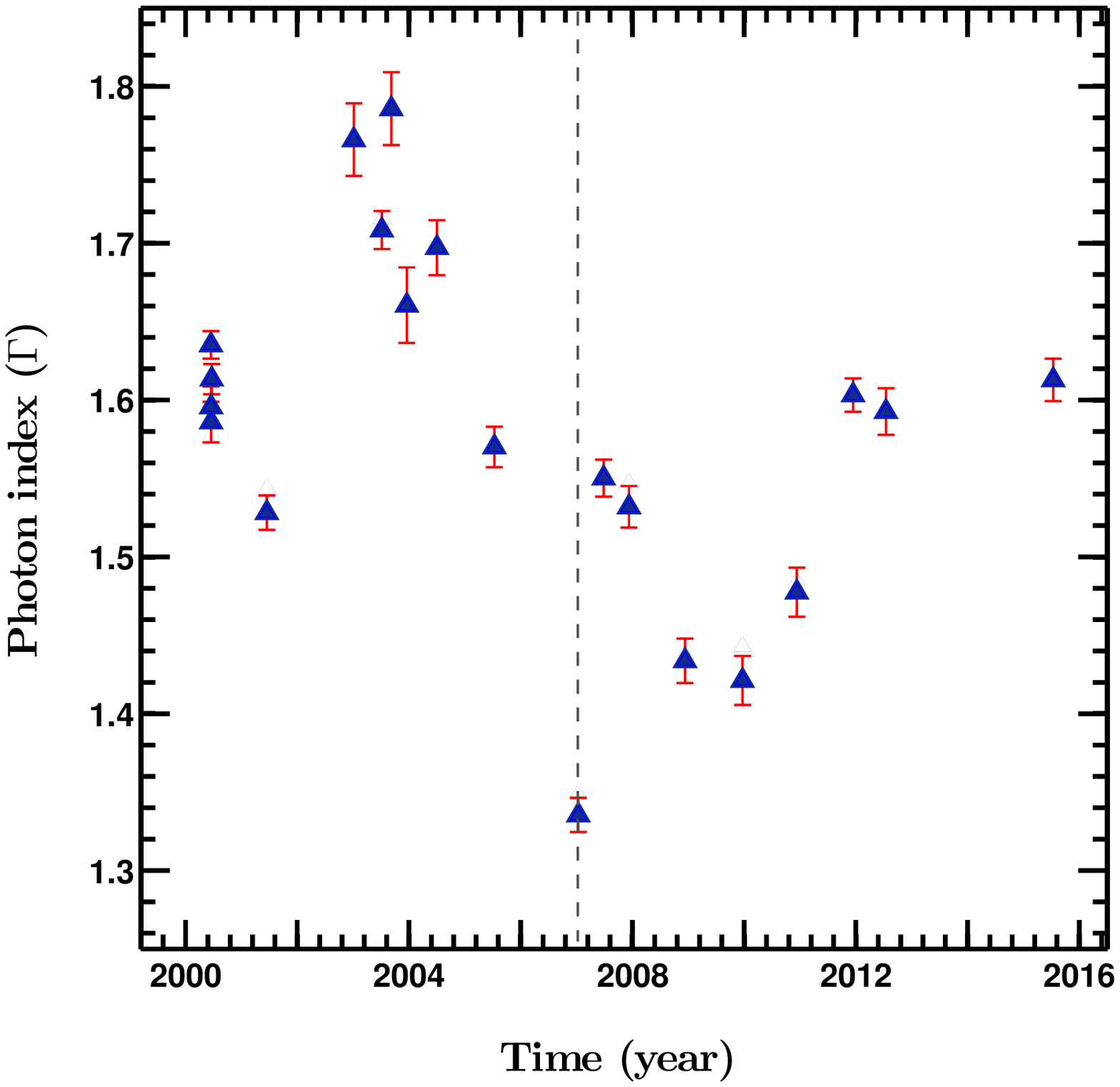, scale=0.5}
\caption{Evolution of the spectral index from June 2000 to July 2015, showing a significant variation of $\Delta\Gamma$ = 0.44. The dotted line emphasizes the flaring period.}
\label{fig9}
\vspace*{.4cm}
\end{figure}

\section {SPECTRAL EVOLUTION AND CORRELATIONS}

\begin{figure*}      
\centering
\mbox{\subfloat{\includegraphics[width=3.55in]{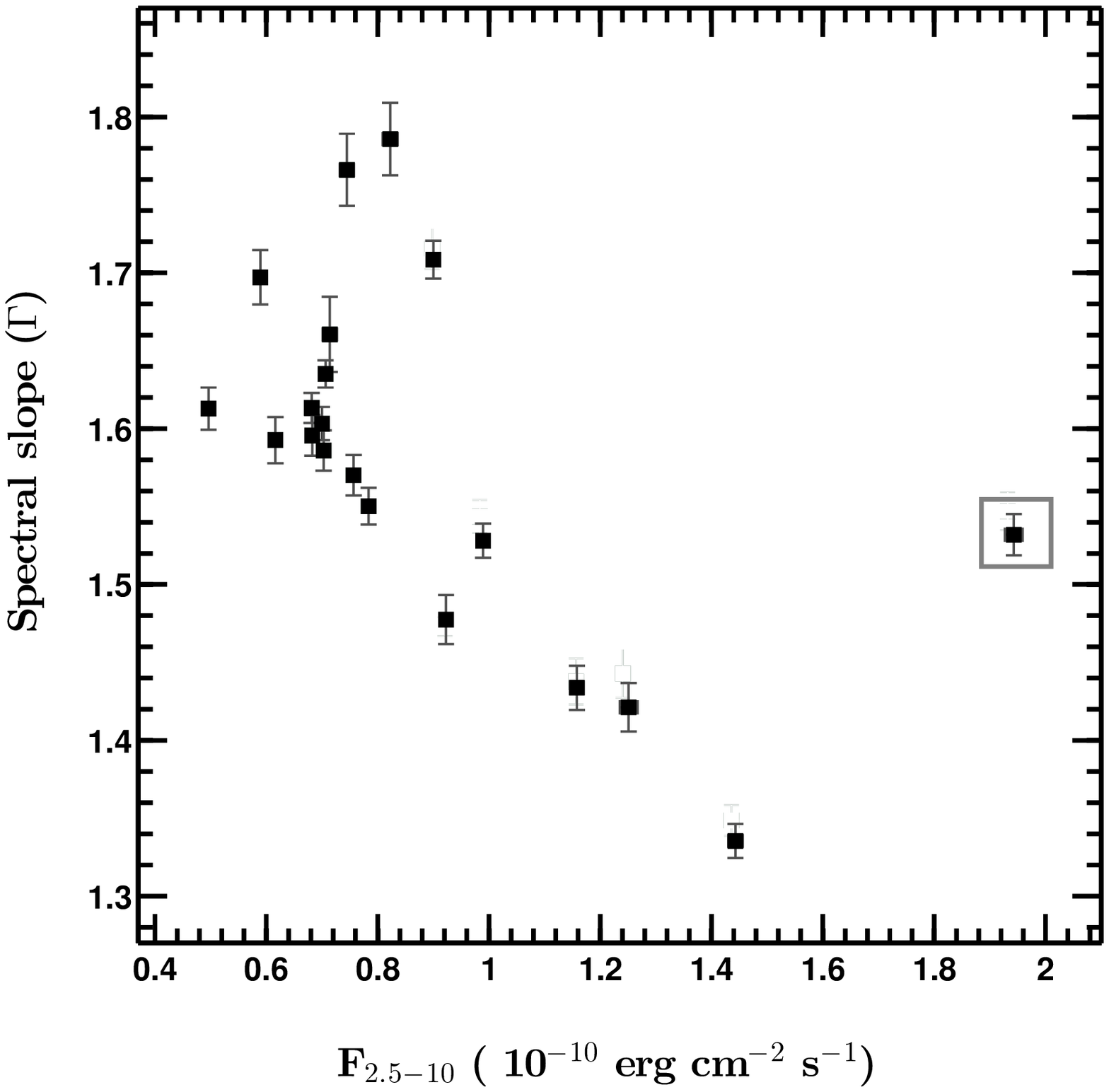}}\quad
\subfloat{\includegraphics[width=3.55in]{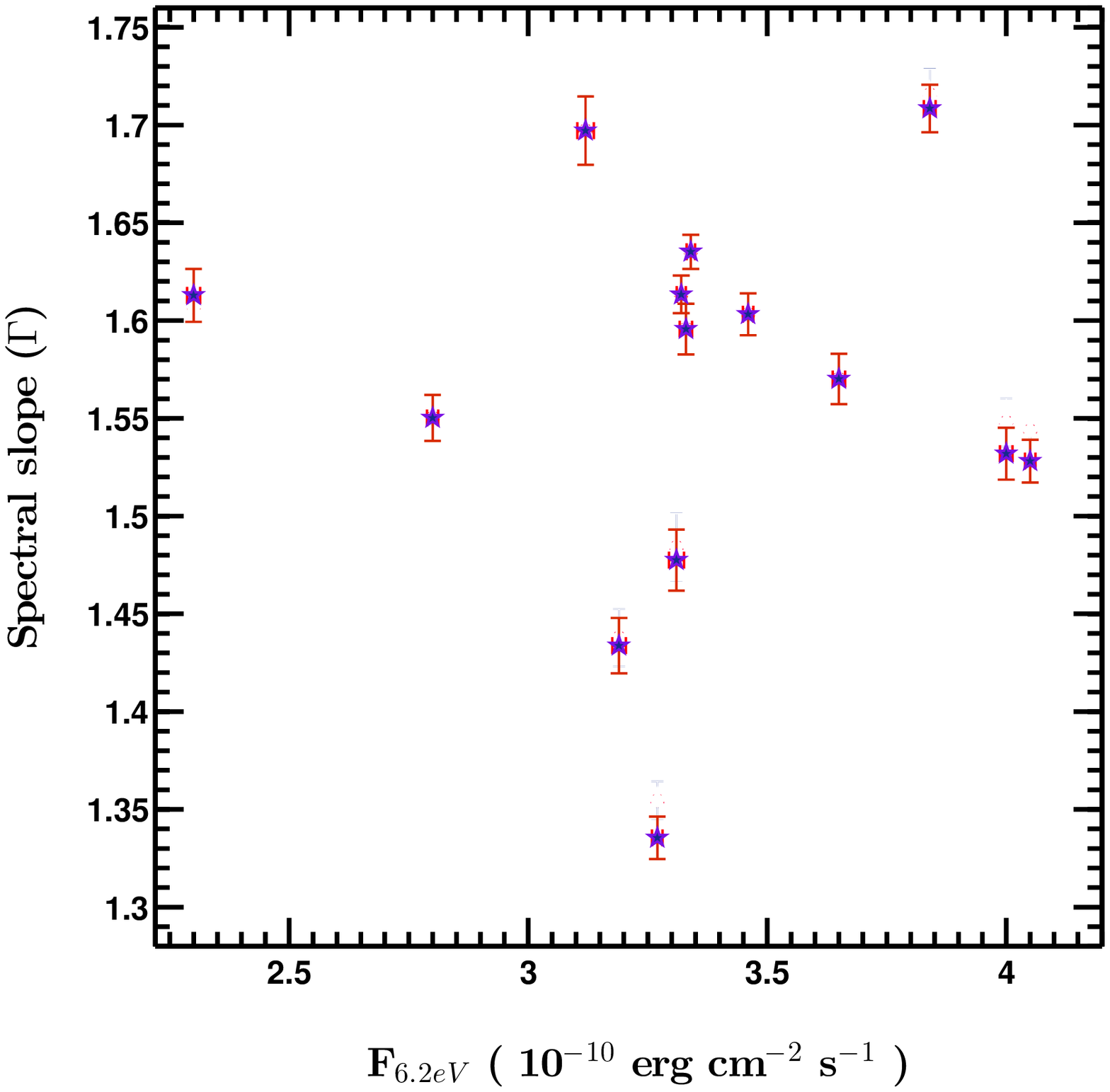} }}
\caption{Left: A strong anti-correlation  between the 2.5--10 keV flux and X-ray spectral slope over that band is present, except for the  rightmost point inside the small box that corresponds to the 2007 X-ray flare. 
Right: X-ray spectral slope vs.\ ultraviolet flux in the 6.2 eV band.}
\vspace*{0.7cm}
\end{figure*}

\subsection {Variation of the power-law continuum}

Before studying the evolution of the spectra, we investigate the emission states of the source at the time of these {\it XMM-Newton} observations. Multi-wavelength monitoring of the source in June 2004 revealed a historic minimum in the sub-millimeter, infrared and X-ray bands (T{\"u}rler et al.\ 2006). During the June 2003 XMM observation, the source reached a historically softest state in the X-ray band as reported in Chernyakova et al.\ (2007); they also mentioned that the source evolved towards a softer X-ray state. The source was in a high state in the hard X-ray band (18--60 keV) detected by the AGILE satellite during December 2007 and January 2008 (Pacciani et al.\ 2009). Several consecutive X-ray flares were observed in the source between 2005 to 2012 with sequential high and low states with {\it RXTE-PCA} (Esposito et al.\ 2015), but only one {\it XMM-Newton} observation taken within this period (in December 2007) coincides with the occurrence of a small flare in the source.

From the literature (Page et al.\ 2004; Chernyakova et al.\ 2007; Esposito et al.\ 2015) we find that, except for the flaring period mentioned above, all the {\it XMM-Newton} observations we are analyzing were carried out during a low/quiescent state of 3C 273. This gives us an excellent opportunity to study the long term spectral behaviour of this FSRQ in the low state.  

 In Figure 3, we have plotted the temporal variation of X-ray fluxes extracted from individual spectral fits along with the ultraviolet fluxes from simultaneous OM observations. Between mid-2000 and mid-2015 the 2.5--10 keV flux changes by a factor of more than three, being $4.98\times 10^{-11}$ erg cm$^{-2}$ s$^{-1}$ as the lowest flux (July 2015) and $1.9\times 10^{-10}$ erg cm$^{-2}$ s$^{-1}$ during the flaring period. The 2.5--10 keV X-ray flux in 2015 calculated here has a smaller value than the flux, $5.89\times 10^{-11}$ erg cm$^{-2}$ s$^{-1}$,  found at the historically faintest period, June 2004. It can be seen from the figure that during 2015 the UV band flux is also the lowest in comparison to other epochs. Thus we can say that 3C 273 reached a new minimum flux state in 2015 which is even fainter than the historically low flux state during 2003--2004. From the figure we can also see that the flaring period indicates the highest flux state in both X-ray and UV bands ($1.9\times 10^{-10}$ erg cm$^{-2}$ s$^{-1}$ and $4.0\times 10^{-10}$ erg cm$^{-2}$ s$^{-1}$, respectively).  

We recall that the overall shapes of the X-ray spectra reflect the underlying X-ray emission mechanisms but variations of the spectral slope can indicate changes in the geometry of the X-ray producing region. We will discuss this in some detail in section 5. The temporal spectral variation of the source from 2000 to 2015 is shown in Figure 4.  During this period the spectral slope varies significantly, with  $\Gamma$ ranging between 1.35 and 1.79.  In the energy band 2.5--10 keV, we found a strong anti-correlation between the spectral slope and the flux, i.e., a ``harder-when-brighter trend", in the sense that the spectra get harder with increasing flux (Figure 5, left). Excluding the data point taken during the flare, the Pearson correlation between these two parameters is $R = 0.72 $ with 0.9995  probability of correlation.  We also checked if there was any relation between the X-ray spectral slope and the UV flux but find that these quantities do not appear to be related (Figure 5, right).   In addition, the normalization to the power-law, $N$, is not related to the photon index (Figure 6, left) but is, unsurprisingly, connected to the flux, with a correlation coefficient of 0.85  (including the flaring point), as depicted in Figure 6 (right).  In these plots the flaring point is usually an outlier, and to make that clear, it is shown inside a small square box.

\begin{figure*}      
\centering
\mbox{\subfloat{\includegraphics[width=3.55in]{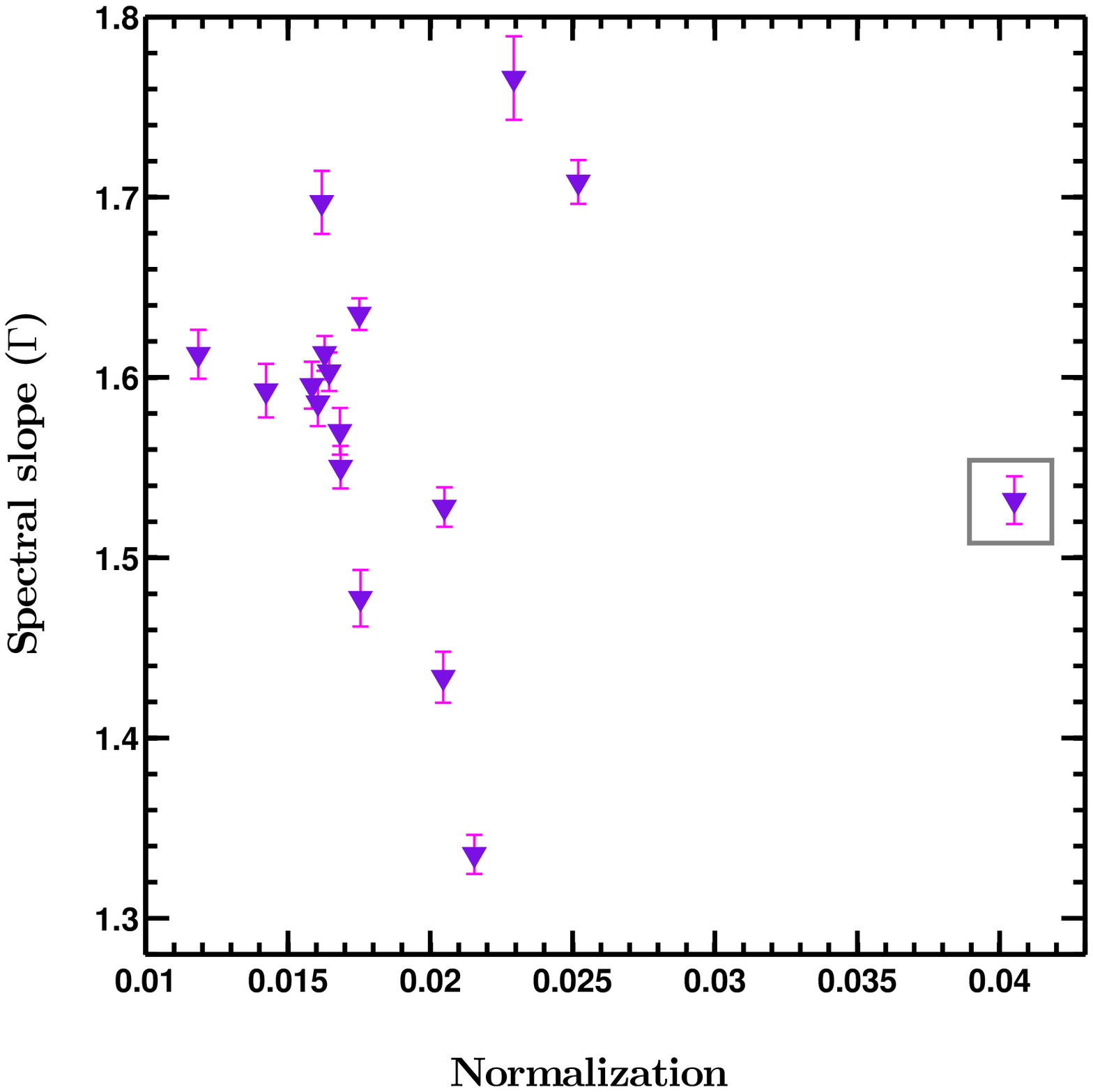}}\quad
\subfloat{\includegraphics[width=3.55in]{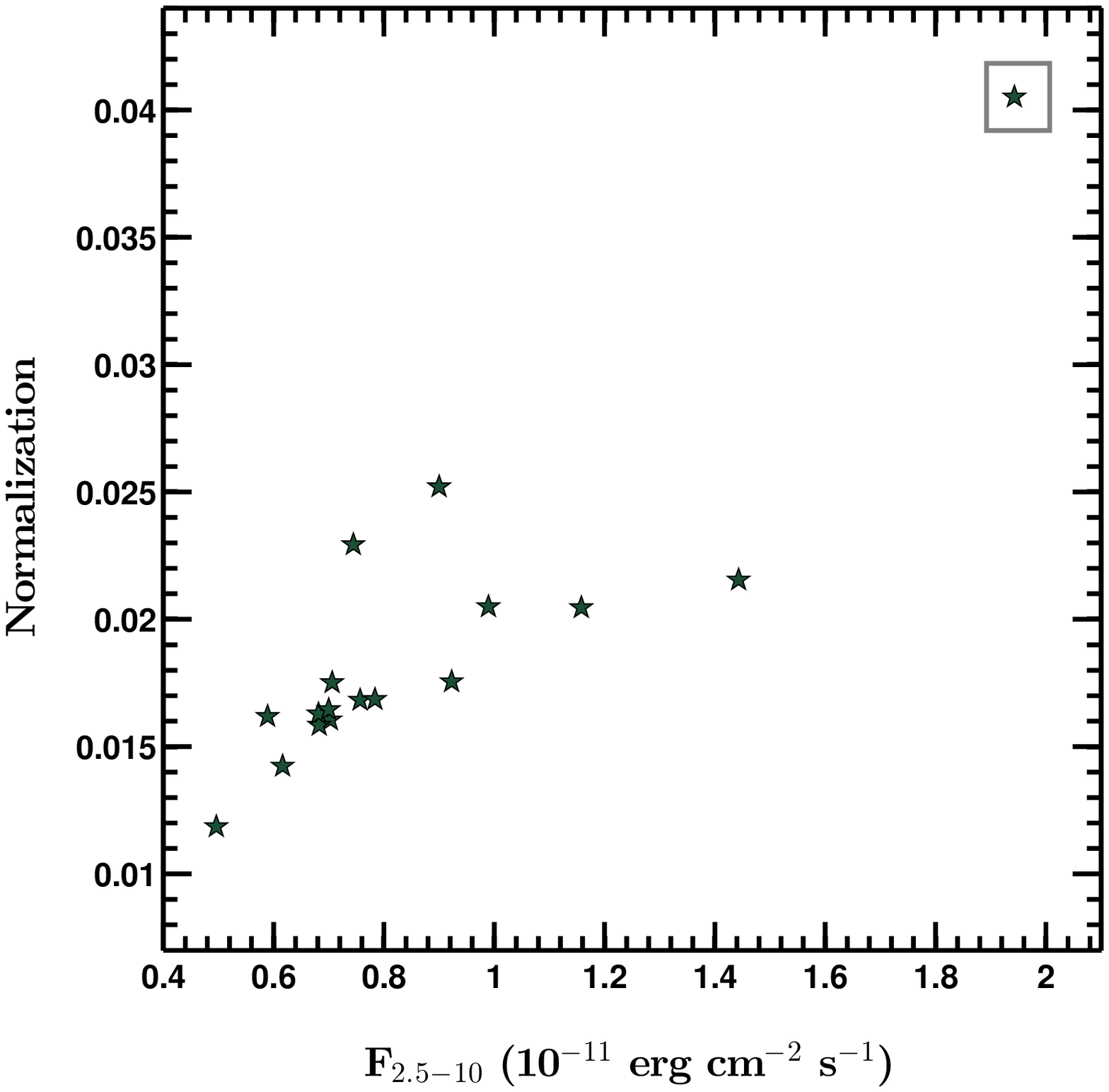} }}
\caption{Left:  X-ray spectral index as function of the normalization used for the fitted power-law. Right: Normalization vs.\ spectral flux. X-axis error bars in both figures are smaller than the point size. Normalization is in units of photons keV$^{-1}$cm$^{-2}$s$^{-1}$.}
\vspace*{0.7cm}
\end{figure*}

\begin{figure*}      
\centering
\mbox{\subfloat{\includegraphics[width=3.5in]{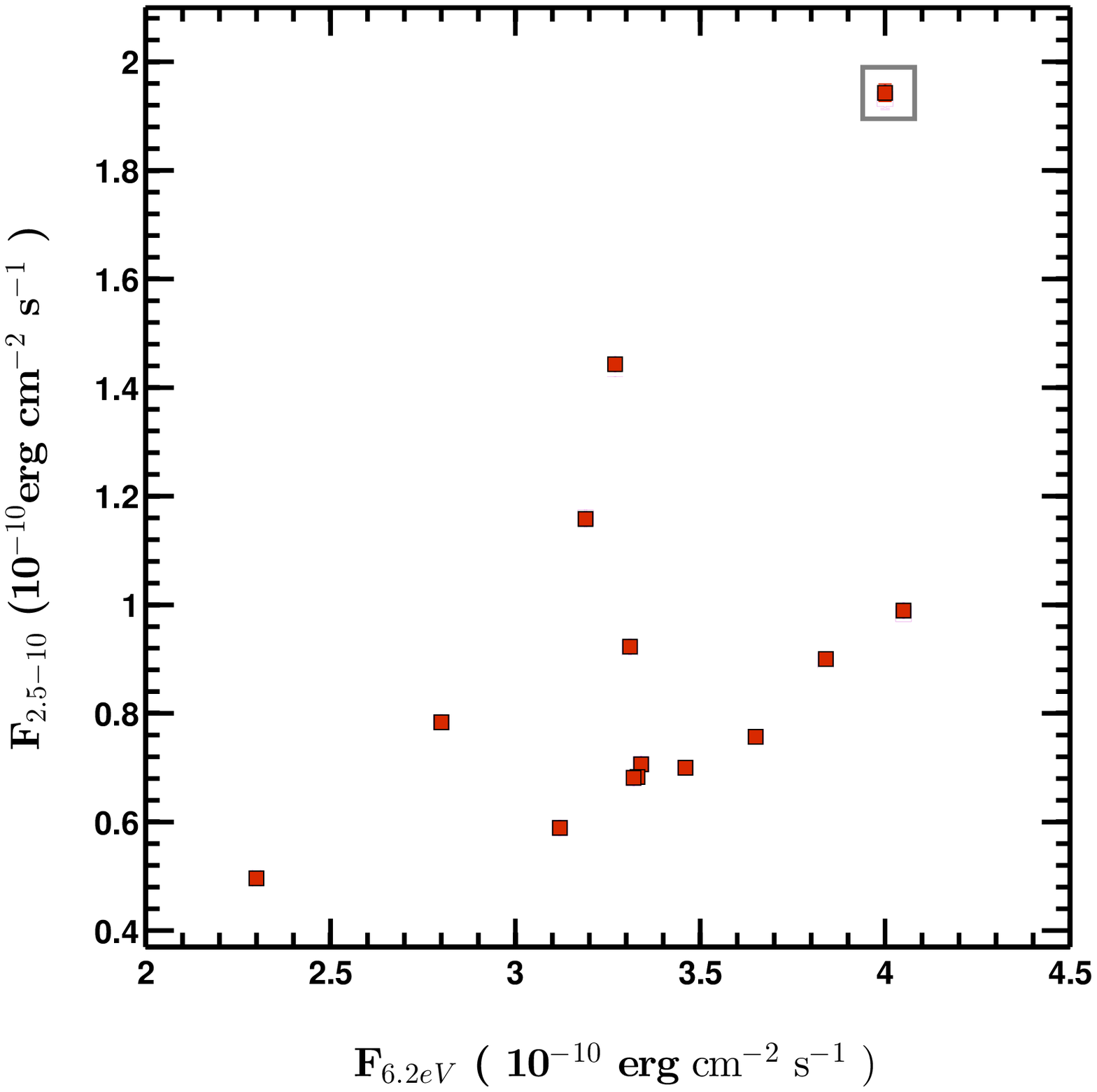}}\quad
\subfloat{\includegraphics[width=3.5in]{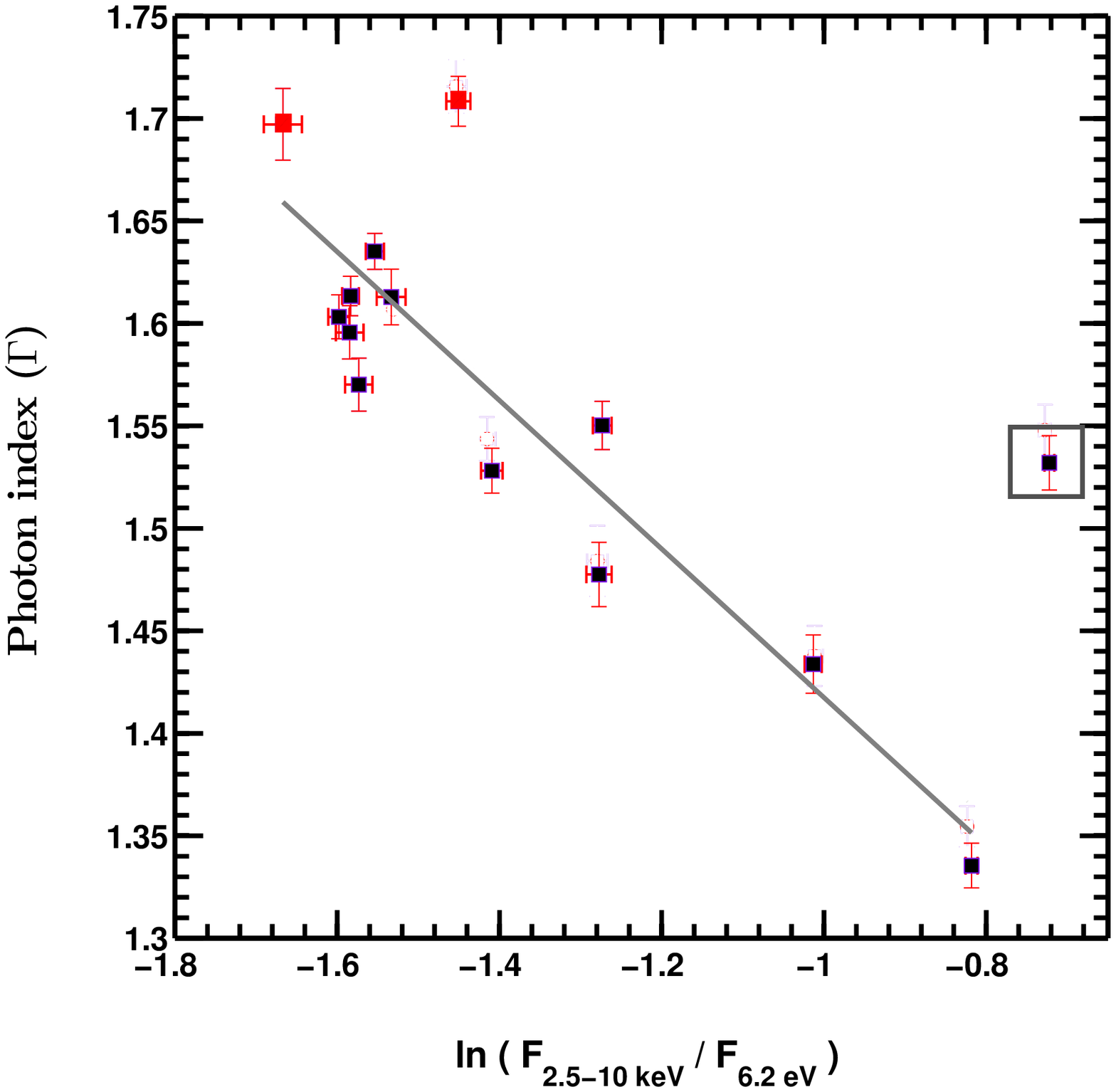} }}
\caption{Left: X-ray flux in the energy range 2.5--10 keV vs. ultraviolet flux at energy 6.2 eV. Error bars are smaller than the symbol size.  Right: The 2.5--10 keV spectral index as a function of the logarithm of the ratio of the 2.5--10 keV to 6.2 eV UV flux shows a significant anti-correlation.}
\vspace*{0.7cm}
\end{figure*}

\subsection {{\bf Possible} correlation between X-ray and UV emission}

A long standing question regarding the relevance of Compton scattering of UV photons to X-ray energy is still unresolved in 3C 273. 
The X-ray emission from AGNs can result either from a continuation of the synchrotron spectrum or from inverse Compton emission from low energy seed photons scattered off a thermally hot distribution of electrons (corona-like structure) situated above the accretion disc or from inverse Compton scattering off the relativistic electrons in the jet. Since 3C 273 is a borderline object and shows both Seyfert-like and blazar-like emission, an  inverse Compton mechanism has been introduced to explain X-ray emission from time to time (Sunyaev \& Titarchuk 1980; Bezler et al.\ 1984; Leach et al.\ 1995; Grandi \& Palumbo 2004). Because of observational constraints and the lack of simultaneous observations in these bands, this thermal inverse Compton explanation requires additional evidence, as significant proof of this point may expand our understanding of AGN emission mechanisms.  Fortunately, {\it XMM-Newton}, with its Optical Monitor (OM)  attached to the X-ray telescope, can observe a source in both X-ray as well as in multiple optical/UV bands at the same time.  Taking advantage of this facility, we can try to understand the origin of the X-ray emission for this source in the low state. 

Walter $\&$ Courvoisier (1992) for the first time found a correlation between X-ray (2--10 keV) and UV (10 eV) emission in 3C 273 with Ginga, {\it EXOSAT} and {\it IUE} quasi-simultaneous X-ray and UV observations taken during 1984 to 1988. They studied the 2--10 keV spectral energy index vs.\ logarithm of the ratio of the 2--10 keV to UV count rates and found an linear anti-correlation between them. This strongly supports the thermal inverse Comptonization model as being responsible for this portion of the  X-ray emission in the source. But later,  Chernyakova et al.\ (2007)  did a similar analysis with some early {\it XMM-Newton} observations (2003--2005) where they did not find any correlation.

Following their work, we have reexamined this relation with 15 years of {\it XMM-Newton} data, obtained simultaneously for both X-ray and UV bands.  In the left panel of Figure 7 we plot the X-ray flux against the UV flux and note that though there is a great deal of scatter there appears to be a weak linear relation between them with correlation coefficient of 0.50 (if the flaring point is included). In the right panel of Figure 7 we have plotted the 2.5--10 keV spectral index as a function of logarithm of the ratio of the 2.5--10 keV to UV flux, as extracted from UVW2 filter. There we clearly see an  anti-correlation, as was found in Walter $\&$ Courvoisier (1992).  
Recall that nearly all of these observations were taken during the low state of the 3C 273. The solitary  exceptional point at the extreme right corresponds to the 2007 X-ray flare. Excluding this point, a least square fitting to the rest of the points gives a correlation coefficient of 0.89 with a slope of -0.362$\pm$0.023. 

However, it is possible that this observed anti-correlation between $\Gamma$ and the ratio of the X-ray to UV fluxes (Fig.\ 7, right) is not a physically relevant one, in that the distribution of points in the plot of $\Gamma$ against UV flux (Fig.\ 5, right) could be considered to be dominated by two distinct groups containing all but one of the points between them.  One vertical band, composed of the middling fluxes ($3.1 - 3.5\times 10^{-10}$ erg cm$^{-2}$ s$^{-1}$) shows a wide range of $\Gamma$ values, while one horizontal band ($1.52\leq \Gamma \leq1.64$) exhibits a wide range of fluxes.
Given the anti-correlation between $\Gamma$ and $F_{\rm X}$, the vertical group automatically produces an apparent anti-correlation between $\Gamma$ and log($F_{\rm X}/F_{\rm UV}$) since one is just dividing by an essentially constant $F_{\rm UV}$.




\section{ DISCUSSION }

\begin{figure*}      
\centering
\mbox{\subfloat{\includegraphics[width=3.55in]{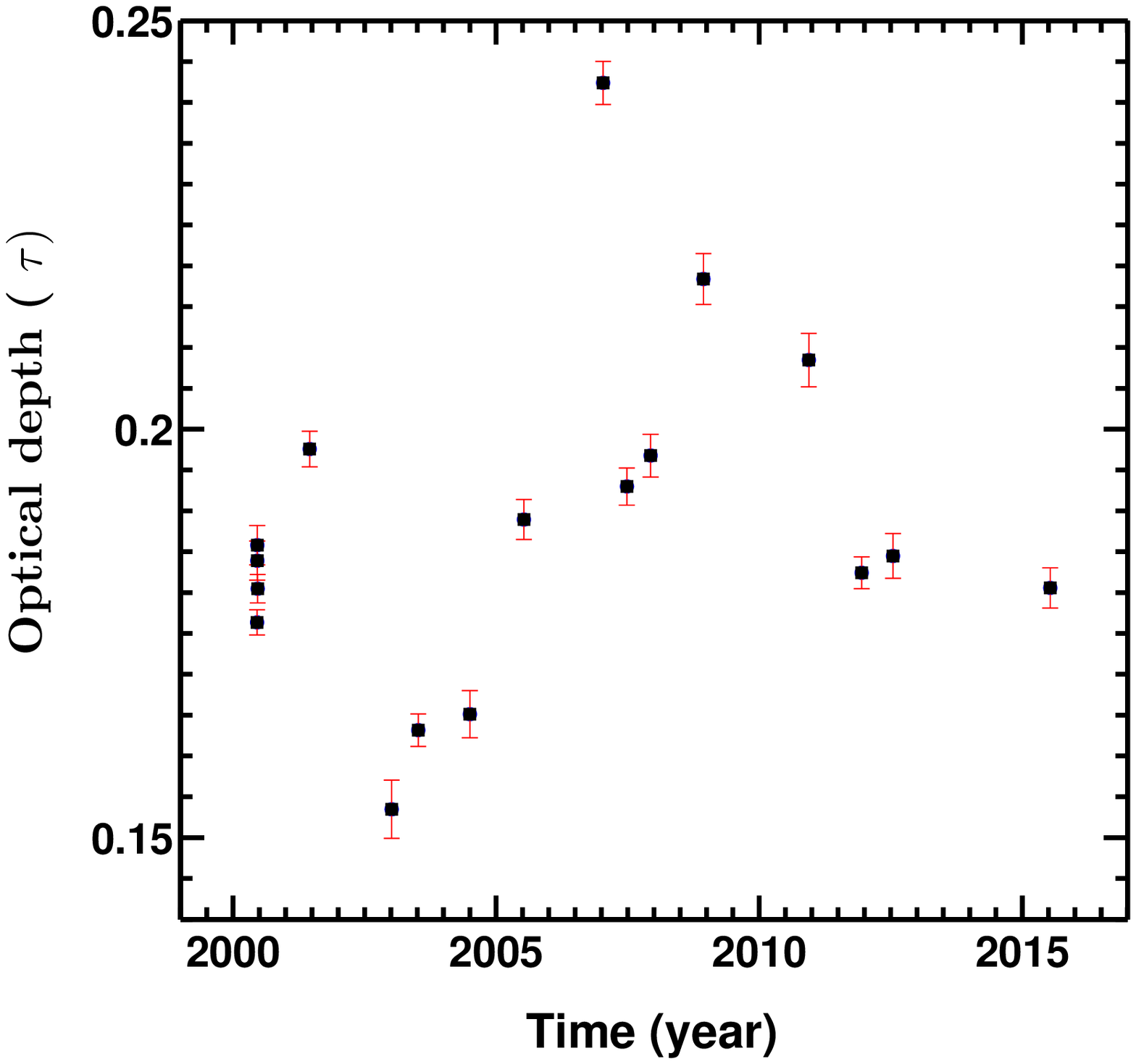}}\quad
\subfloat{\includegraphics[width=3.51in, height=3.41in]{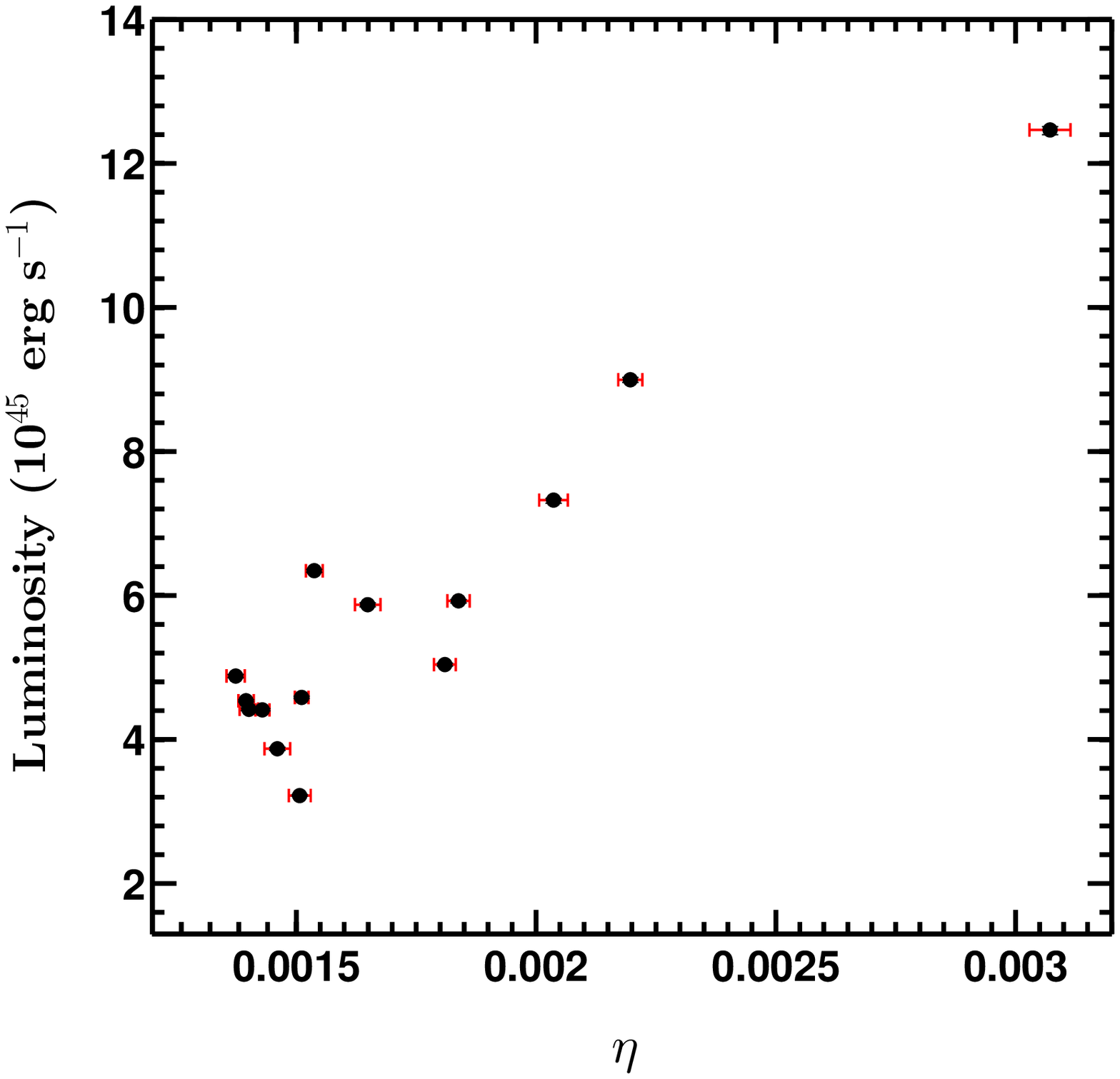} }}
\caption{Left: Variation of optical depth of the comptonizing medium at different epochs. Right: X-ray luminosity of the source vs.\ covering factor of the Comptonizing medium. Vertical error bars are smaller than the symbol size.}
\vspace*{0.7cm}
\end{figure*}

We have shown in section 3 that almost all the spectra of 3C 273 between 2.5--10 keV energy range, as observed during 2000 to 2015 by {\it XMM-Newton} are well described by a simple power-law model. As is typical for AGNs where sufficient data is available, spectral variation can be seen in 3C 273. In section 4, we presented the temporal variation of X-ray spectral slope of 3C 273 and its relation with several spectral parameters. 

 In this work we have found a very clear anti-correlation between the slope of the X-ray spectra and the flux of 3C 273, which means the spectra flatten with increasing flux, as is clearly visible from Figure 5. Surprisingly, this is totally opposite what reported in Kataoka et al.\ (2002), where the spectra became steeper when the source is brighter, similar to that typically observed in Seyfert galaxies. The critical point  is that nearly all the observations used for our study were taken when the source is in a low or quiescent period. The exception to this relation occurs during the December 2007 X-ray flare reported in Esposito et al.\ (2015). 

In most of the previous studies of this source any relation between the slope and flux  was absent (Turner et al.\ 1990; Page et al.\ 2004; Chernyakova et al.\ 2007; Soldi et al.\ 2008). However, a softer when brighter pattern of the 2--10 keV X-ray spectra was reported in Figure 6 of Kataoka et al.\ (2002) with {\it RXTE} observations made during 1999--2000, while this relation was absent with the 1996--1997 observations when the source was in a relatively lower flux state than in 1999--2000. The positive slope--flux relation reported by Kataoka et al.\ (2002) was interpreted by them in the context of non-beamed, thermal emission from the accretion disc and reprocessed emission from the thermal corona. So the different circumstances under which the slope-flux correlation is present or not require further investigation in future observations.  
An anti-correlation between slope and flux was reported in Madsen et al.\ (2015) with 244 ks continuous {\it NuSTAR} exposure of 3C 273 made in July 2012 in the energy range 3--10 keV, where they have postulated that this type of variation is driven by the jet emission while the spectra harden with an increasing jet contribution to the flux. But our results do not concur with their interpretation since we have found the anti-correlation during the low states of the source when the emission from the jet is almost certainly at a minimum. Moreover, the indication of a connection between optical/UV and X-ray bands seen in our study, support a different scenario where the X-ray photons are emitted through inverse Compton scattering of UV photons from a thermal corona (see section 4.2 and below). 

Hayashida et al.\ (2015) reported a similar spectral variability; harder-when-brighter (for soft spectra, 0.5--5 keV), in 3C 279 during a flaring state of the source with {\it SWIFT-XRT} data. It has been extensively reported and believed that the emission mechanisms and hidden physical scenario in 3C 279 and 3C 273 are similar and both belong to the same category (border-line objects). But the presence of a slope-flux anti-correlation in different states of the sources make for a more complicated emission behaviour. In general, a harder-when-brighter trend is seen in the X-ray spectra of high frequency peaked blazars or in flaring blazars where the SSC emissions from the jet play dominant role (Krawczynski et al.\ 2004; Gliozzi et al.\ 2006; Emmanoulopoulos et al.\ 2012).  

The X-ray emission of FSRQs is quite complex and interesting, and the  physical mechanism responsible for the emission is still not understood very well. In radio quiet AGNs, particularly Seyfert galaxies, the slopes of the X-ray spectra are linearly correlated with the 2--10 keV X-ray flux i.e., the spectra steepen with increasing fluxes (Sobolewska \& Papadakis 2009; Caballero-Garcia et al.\ 2012, and references therein). This softer-when-brighter behaviour in Seyfert galaxies in the 2--10 keV band has been explained through different spectral variability models. Such spectral variability might be totally intrinsic and may arise from variation of accretion rate in the source, i.e., softer X-ray spectra correspond to higher accretion rates (Shemmer et al.\ 2006; Sobolewska \& Papadakis 2009). Moreover, there could be a X-ray absorber with variable ionization state which is directly related to the intrinsic variation of the power-law continuum or with a variable column density or covering fraction and thus with a variable opacity along the line of sight, which absorbs X-rays from a constant intrinsic continuum (Sobolewska \& Papadakis 2009; Turner et al.\ 2007). Another interpretation for softer-when-brighter spectra is that the spectral variation is a result of combined contributions from a power-law continuum that is highly variable in flux but with a constant spectral shape and a constant reflection component (Taylor et al.\ 2003; Ponti et al.\ 2006).  Or this trend may occur due to superposition of a constant reflection component and a variable continuum which is variable in both flux and shape (Sobolewska \& Papaderakis 2009).

It has been suggested that the observed softer-when-brighter or harder-when-brighter trend is related to the accretion rate of the system. The harder-when-brighter correlation was reported in Gu \& Cao (2009) for a sample of low luminosity AGNs (LLAGNs) and interpreted in the context of advection-dominated accretion flows (ADAFs), where the X-ray photons are result of Comptonization of seed photons in the ADAF. A similar trend was observed in a sample of LINERs by Younes et al.\ (2011). Wu \& Gu (2008) reported that in case of black hole X-ray binaries (BHXRBs) the harder-when-brighter trend is observed when the BH accretion rate is below a critical value. Above this critical accretion rate a softer-when-brighter trend is observed. A similar behaviour has been observed in AGNs with different studies (Gu \& Cao 2009; Younes et al.\ 2011; Connolly et al.\ 2016), supporting the accretion rate dependence of spectral trend as observed in the case of BHXRBs. Another possible explanation of this behaviour is an outflowing hot corona situated above an untruncated accretion disc, where the seed photons for Compton scattering come from the disc (Sobolewska et al.\ 2011). 

The slope-flux correlations observed in 3C 273 in all previous studies are based on short term spectral studies of the source. From our result and earlier studies, we conclude that the short term positive slope-flux relation reported in Kataoka et al.\ (2002) is found only in high states of the source while the negative slope-flux relation is seen during low states of 3C 273.  So the harder-when-brighter trend observed here for 3C 273 during its low states might be the result of smaller accretion rates.  

The higher sensitivity of {\it XMM-Newton} EPIC-pn data, allowing for an excellent signal to noise ratio, gives us more precise spectral parameters than do all the previous studies.  In addition to this, the ability of {\it XMM-Newton} to take simultaneous X-ray and optical observations of a target gives us more useful results.  Considering the data as a whole, it seems fair to conclude that the appearance, or not, of a relation between the X-ray slope and flux depends on the source state. We can say that the spectral slope is related to the continuum flux during low states, when the contribution to the continuum from the jet emission is minimum. It is worth noting that the apparent absence of this relationship in some earlier work may also be due to the small number of spectra available or because of the poorer quality of data.  

In section 4.1, we examined the relation between UV (6.2 eV) and X-ray emission through simultaneous observations in these bands.  When the UV fluxes are plotted against X-ray fluxes (left panel, Fig.\ 7),  a weak correlation between them was observed, but we saw a somewhat stronger correlation between X-ray spectral index and the ratio of the X-ray to UV fluxes during the generally low state (right panel, Fig.\ 7). The observations carried out during the historical low state of the source  in 2003 (upper left points in that plot) show greater deviations from the line representing the correlation 
and the observation taken in December 2007 during flaring activity does not follow this pattern.  In a previous study such X-ray flares were argued to be counterparts of infrared/optical flares with a lag of few months (Walter $\&$ Courvoisier 1992) and VLBI radio observations found that times of X-ray flares coincided with the birth of new radio knots (Baath et al.\ 1991).  Considering these observations, we checked the literature for such connections to the 2007 X-ray flare in the source. The mm flux reached its minimum in 2004, and since about 2006 this flux has shown high variability (Trippe et al.\ 2011). However, in this case no counterparts of this X-ray flare were noted in either radio or optical bands; moreover, we could find no report on the birth of any radio knot related to this 2007 ``orphan'' X-ray flare.  

If physical, the observed anti-correlation between the hard X-ray spectral slope and the ratio between X-ray and UV fluxes can be interpreted in the context of coronal emission.  The presence of a hot corona located in the inner region of the central engine of AGN is well accepted and  widely used to interpret the hard X-ray emission in AGN through multiple up-scattering of soft seed photons by the thermal gas (Haardt \& Maraschi 1991; Nandra et al. 2000; Nandra \& Papadakis 2001).

It is well known that the X-ray emission in 3C 273 can be disengaged from emission at lower frequencies (radio to optical) which are certainly dominated by the synchrotron emission from the jet. 
A picture of thermal inverse Compton scattering from a corona has been widely applied to interpret the X-ray emission process in the lower energy bands in this FSRQ  (Walter \& Courvoisier 1992; McHardy et al. 1999; Page et al. 2004).  If we consider that inverse Compton scattering is responsible for most of the X-ray emission in 3C 273 then the presence or absence of correlations between slope and flux must be dependent on the variations of the physical parameters of the Comptonizing medium.  In particular, the variation of the X-ray spectral slope must be related to the variation of physical conditions in the medium scattering those X-rays. Most of the observations used in this study belong to quiescent states of the FSRQ, i.e., when jet emission is minimal, hence it is quite possible that the X-ray photons  originate from inverse Compton scattering of seed photons by a hot electron gas. If the Comptonization is purely thermal then the slope will be governed by both the temperature and Thomson optical depth of the electron distribution in the medium. Since the number of incident soft seed photons to the number of emerging high energy photons ratio is directly proportional to the optical depth of the scattering medium, this must be tightly tied to the slope of the power-law describing the Compton emission.

For inverse Compton process in an optically thin electron gas, the rate of production of X-ray photons from UV photons is given by the relation (Walter \& Courvoisier 1990)  
\begin{equation}
N_{X} = \eta (1-e^{-\tau}) N_{UV},
\end{equation}
where $N_{X}$ and $N_{UV}$ are the emission rates of inverse Compton processes and soft seed photons from the BBB, respectively; $\eta$ is the covering factor of the thermal gas surrounding the source of UV seed photons; and $\tau$ is its optical depth of the scattering medium. The above equation can be simplified to
\begin{equation}
N_{X} \thickapprox \eta \tau N_{UV},
\end{equation}
when $\tau$ is small.

The slope of the power-law spectrum $\Gamma$ triggered by inverse Compton emission is given by Rybicki and Lightman (1979), as
\begin{equation}
\Gamma = - ln(\tau)/ln(A), 
\end{equation}
where $A$ is the fractional energy gain per scattering, or the amplification factor of the Compton scattering.  For mildly relativistic electrons this is given by (Sunyaev \& Titarchuk 1980), 
\begin{equation}
A=16(kT/m_{e}c^{2})^{2}.
\end{equation}

If we consider that the X-ray spectrum is generated mainly by the inverse Compton emission from a relativistic hot electron gas triggered by the low energy ultraviolet seed photons,  a relation between the photon index of the X-ray spectra and the ratio of X-ray flux to the UV flux can be established as (Walter \& Courvoisier 1990, 1992):
\begin{equation}
\Gamma = - \frac{{\rm ln}(F_{2.5-10 keV} / F_{6.2 eV})}{{\rm ln} (A)} + f(\eta,A).
\end{equation}

Here, the correction term $f(\eta,A)$, is negligible for low optical depth.  
From the above equations we can find the gain factor, {\it A}, from the slope of the spectral index vs.\ flux ratio when the X-ray flux and spectral variability correlation is caused by variation of optical depth.  Once {\it A} is known, then the temperature, {\it T} of the Comptonizing medium can be calculated  from Eq.\ (4). The optical depth $\tau$, and covering factor $\eta$, of the thermal gas can be estimated from Eq.\ (2) and Eq.\ (5), respectively. From the anti-correlation found in Fig.\ 7, which is nothing but $\Delta \Gamma /\Delta {\rm ln}(F_{2.5-10 keV} / F_{6.2 eV})$, the other parameters can be estimated. The estimated value of the temperature (kT) of the thermal gas is $5.4 \times 10^{2}$ keV, and, confirming our assumption of small optical depth, the $\tau$ values lie in the range 0.18--0.24. We next examine the variation of optical depth of X-ray emitting medium with time and display it in the left panel of Figure 8.  The optical depth value is highest during the ``flaring state'' in 2007 and has its smallest value in a very low state in 2003. If we consider the the X-ray source as isotropic then the value of covering factor $\eta$ is found to be in the range $\thickapprox$ 0.0014--0.0031 which is substantially lower than the value of $\thickapprox$ 0.02 estimated by Courvoisier \& Camenzind (1989).  The covering factor is apparently roughly linearly related to the luminosity of the source, as shown in the right panel of  Figure 8.

The broadband X-ray spectrum of the source may be described in terms of  a combination of AGN components comprising coronal emission and disk reflection and a jet component which shows a transition at around 20 keV (Grandi \& Palumbo 2004; Esposito et al.\ 2015; Madsen et al.\ 2015). Fitting the {\it NuSTAR} spectra (3--78 keV) with a CompTT model, to probe a  physical model for the coronal continuum without considering the jet component, gives a plasma temperature of, kT$_{e}$ = 247$^{+69}_{-64}$ keV and a optical depth $\tau$ = 0.15$^{+0.08}_{-0.04}$. However, inclusion of jet emission in this model reduces the plasma temperature to 12.0$\pm$ 0.3 keV while $\tau$ increases to 2.77$\pm$ 0.06 (Madsen et al.\ 2015). 
The correlated variation of blackbody and Seyfert-like (power-law) fluxes observed by Grandi \& Palumbo (2004) with {\it BeppoSAX} data can be understood in terms of a geometrically thin accretion disc which is responsible for the black-body emission, while  inverse Compton scattering in the disc corona produces the high energy Seyfert-like component. Another study done by Esposito et al.\ (2015) suggested that SSC emission from the relativistically beamed jet is the main source of high energy photons above 100 MeV, while the X-ray photons are most likely inverse Compton process occurring in a thermal corona near the central engine. They suggested that the variations between different flaring states of the source are governed by changes in electron Lorentz factors.

These results possibly can be explained in the context of accretion disc winds.  Accretion discs are associated with intense and twisted magnetic fields and while the topology of the magnetic lines is still not understood properly, various simple models provide reasonable places to start.  Much of the energy generated in the disc can be dissipated either as a wind or as a relativistic jet in the form of bulk outflow (e.g.\ Blandford $\&$ Payne 1982).  Livio et al.\ (2003) proposed a model for a disk-jet connection in X-ray transients and AGNs, wherein  the inner accretion disc  has two states.  In one state, the energy emitted within the disc may dissipate locally in the form of a disc wind into the low-density regions above and below the disc to produce a  corona-like structure. In the other state, a huge amount of energy emerges from the inner part of the disk as collimated relativistic jets propelled by large scale magnetic fields. In the first case of the inner disc wind, magnetic forces  generate shocks in the wind 
fields, which thermalize both protons and electrons.  The electrons are further energized as a result of interactions with the thermal protons. These highly energetic electrons are later cooled off by inverse Comptonization of seed UV photons coming from the inner disc or from the jet.  Within this framework, in the low state of 3C 273 the inner accretion disc is in the first state, which is insufficient to project the material and radiation in the form of collimated jet. Then the  hot electron gas is carried out by disc winds to a  region where they eventually up-scatter the UV seed photons into X-rays.   

\section{CONCLUSIONS}

In this paper, we have presented a new analysis of simultaneous X-ray and ultraviolet observations of 3C 273, taken with {\it XMM-Newton} during 2000--2015. The key conclusions of this work are:\\

--- Almost all of the 20 quality X-ray spectra in the  2.5--10 keV energy range are well described by a single power-law model with Galactic absorption due to neutral hydrogen along the line of sight to 3C 273.  All but one of these measurements were made while 3C 273 was in an overall low state.\\ 

--- A new lowest X-ray flux state of 3C 273 was observed in July 2015. During this period the 2.5--10 keV flux ($4.98\times 10^{-11}$ erg cm$^{-2}$ s$^{-1}$) is even lower than the flux observed ($5.89\times 10^{-11}$ erg cm$^{-2}$ s$^{-1}$) at the historically faintest period (June 2004).\\

--- A weak broad iron emission line was detected at 5.8 keV during a low state of the source in 2003. The line is well represented by a Gaussian profile of equivalent width 47 eV. The width of the line gives an equivalent FWHM value \textgreater 20,000 km s$^{-1}$. Very broad emission line features at energies between 5--7 keV appeared to have been present in 4 other observations, but they were too weak to be well fit.\\

--- The X-ray continuum of 3C 273 in the energy range 2.5--10 keV varies significantly even in the low state of the source. A harder-when-brighter trend of the X-ray spectra is clearly observed, which can be explained either as a result of lower accretion rate in the central engine or through an outflowing hot corona situated above an untruncated accretion disc. \\

--- The possible physical interpretation of the apparent anti-correlation between X-ray spectral slope and different optical/ultraviolet emission fluxes would indicate a nearly co-spatial origin of these bands. if this is the case, it supports the picture that the X-ray continuum arises from the comptonization of BBB photons in a medium of a partially relativistic thermal electron gas in the local environment of the AGN's central engine. 

\section*{ACKNOWLEDGEMENTS}

We thank the anonymous referee for important comments, which helped us to improve the manuscript. NK thankfully acknowledge 
Department of Science and Technology, Government of India for financial support vide reference no. SR/WOS-A/PS-56/2013 under  
Women Scientist Scheme to carry out this work. ACG is partially supported by the Chinese Academy of Sciences (CAS) President's 
International Fellowship Initiative (PIFI) (grant no. 2016VMB073). 

This research is based on observations taken with XMM--Newton, an ESA science mission with instruments and contributions 
directly funded by ESA Member Sates and NASA. This work has made use of the NASA's Astrophysics Data System Abstract Service 
and of the NASA/IPAC Extragalactic Database (NED), which is operated by the Jet Propulsion Laboratory, California Institute 
of Technology, under contract with the National Aeronautics and Space Administration.  

{}
\clearpage
\end{document}